\documentclass[journal]{rmaa}
\usepackage{caption}

\def\fov{FoV }
\def\RRab{RRab }
\def\RRc{RRc }
\def\Mstd{$M_{\mbox{\scriptsize std}}$ }
\def\mins{$m_{\mbox{\scriptsize ins}}$ }
\def\fdif{$f_{\mbox{\scriptsize diff}}$ }
\def\fref{$f_{\mbox{\scriptsize dref}}$ }
\def\sref{$\sigma_{\mbox{\scriptsize ref}}$ }
\def\sdiff{$\sigma_{\mbox{\scriptsize diff}}$ }
\usepackage{xcolor}

\begin{document}

\title{The variable star population in the globular cluster\\ NGC 6934}

% Note the difference betwen `and' and `\and'. The latter appears on a
% line of its own.
  %% note that the `\and' command is necessary to put the `and' on its
  %% own line
\author{
  M. A. Yepez,\altaffilmark{1}
  A. Arellano Ferro,\altaffilmark{1}
  S. Muneer,\altaffilmark{2}
  Sunetra Giridhar\altaffilmark{2}
}

\altaffiltext{1}{Instituto de Astronom\1a, Universidad Nacional Aut\'onoma de
M\'exico, M\'exico.}

\altaffiltext{2}{Indian Institute of Astrophysics, Bangalore, India}

\fulladdresses{
\item M. A. Yepez, A. Arellano Ferro: Instituto de Astronom\1a, Universidad Nacional
Aut\'onoma de
M\'exico, Apdo. Postal 70-264, M\'exico D. F. CP 04510,
M\'exico. (myepez@astro.unam.mx;armando@astro.unam.mx)

}

\shortauthor{Yepez et al.}
\shorttitle{Variable stars in NGC 6934}

\SetYear{2017}
\ReceivedDate{March 2017}
\AcceptedDate{August 2017}

\resumen{Reportamos un an\'alisis de la serie temporal de fotometr\'ia CCD en los
filtros $V$ e $I$ del c\'umulo globular NGC 6934. A trav\'es de la descomposici\'on de
Fourier de las
curvas de luz de las estrellas RR Lyrae, obtuvimos los valores medios de [Fe/H] y
la distancia al c\'umulo; [Fe/H]$_{UVES}$=-1.48$\pm$0.14 y $d$=16.03$\pm$0.42 kpc, y
[Fe/H]$_{UVES}$=-1.43$\pm$0.11 y $d$=15.91$\pm$0.39 kpc, a partir de las
calibraciones de estrellas RRab y RRc respectivamente. Tambi\'en
reportamos valores de la distancia obtenidas con estrellas
SX Phe y SR. Calculamos valores individuales de magnitudes absolutas, radios y masas
para estrellas RR Lyrae individuales. Encontramos 12 nuevas variables: 4 RRab,
3 SX Phe, 2 W Virginis (CW) y 3 semi-regular (SR). La regi\'on inter-modo en la zona
de inestabilidad es
compartida por
las estrellas RRab y RRc. Esta caracter\'istica, observada solamente en algunos
c\'umulos OoI y nunca vista en OoII, se discute en t\'erminos de distribuci\'on de
masa en el ZAHB.}

\abstract{We report an analysis of new $V$ and $I$ CCD time-series photometry of the
globular cluster NGC 6934. Through the Fourier decomposition of the RR Lyrae light
curves, the mean values of [Fe/H] and the distance of the cluster were estimated, we
found; [Fe/H]$_{UVES}$=-1.48$\pm$0.14 and $d$=16.03$\pm$0.42 kpc, and
[Fe/H]$_{UVES}$=-1.43$\pm$0.11 and $d$=15.91$\pm$0.39 kpc, from the calibrations of
RRab and RRc stars respectively. Independent distance estimations from SX Phe and SR
stars are also discussed. Individual absolute magnitudes, radii and masses are
also reported for RR Lyrae stars. We found 12 new variables: 4 RRab, 3 SX Phe, 2 W
Virginis (CW)
and 3 semi-regular (SR). The inter-mode or "either-or" region in the instability strip
is shared by
the
RRab and RRc stars. This characteristic, observed only in some OoI clusters and never
seen in an OoII, is discussed in terms of mass distribution in the ZAHB.}


\maketitle

\section{Introduction}
\label{sec:intro}

Over the recent past, our team has systematically performed CCD photometry of
selected globular clusters aimed to
update the variable star census and to use the light curves of the RR  Lyrae stars to
 estimate the mean distance and metallicity of the cluster in a
homogeneous scale,
and to investigate the dependence of the luminosity of the horizontal branch (HB) on
the metallicity of the stellar system. A summary of the results for a group of some
26 globular clusters has been presented by Arellano Ferro, Bramich \& Giridhar
(2017).

In the present paper we report the results of the analysis of the variable star
population of the globular cluster NGC~6934
(C2031+072 in the IAU nomenclature) ($\alpha = 20^{\mbox{\scriptsize h}}
34^{\mbox{\scriptsize m}} 11.4^{\mbox{\scriptsize s}}$, $\delta =
+07^{\mbox{\scriptsize o}} 24{\mbox{\scriptsize '}} 16.1{\mbox{\scriptsize "}}$,
J2000; $l = 52.10^{\mbox{\scriptsize o}}$, $b = -18.89^{\mbox{\scriptsize o}}$) based
on time-series CCD photometry.
Despite its richness in variable stars, the first variables in this
cluster were discovered rather
late in XXth century. The first 51 variables were reported by Sawyer Hogg (1938).
These stars were further studied and classified by Sawyer Hogg \& Wehlau (1980); fifty
turned out to be RR Lyrae stars and one (V15) was recognized as a
possible irregular variable. Sixty three years passed without the variables in this
cluster
receiving further attention in the literature. Kaluzny, Olech \&  Stanek (2001)
(hereafter KOS01)
performed a time-series CCD photometric study of the cluster and discovered 35 new
variables; 29 RR Lyrae stars, two eclipsing binaries or EW's, two long-period
semi-regular variables or L-SR,
one SX Phe and one unclassified (V85). The identifications of all
these stars can be found in the original chart of Sawyer Hogg (1938) and the small
image cut out's published by KOS01. Their equatorial coordinates are
listed in the Catalogue of Variable Stars in Globular Clusters (CVSGC) of Clement et
al. (2001).

In this paper we describe our observations and data reductions as well as the
transformation to the Johnson-Kron-Cousins photometric system ($\S$ 2), we perform the
identification of known variables and report discovery of a few new ones ($\S$ 3), we
calculate the physical parameters via the Fourier decomposition for RR Lyrae stars
($\S$ 4), highlight the properties of the SX Phe stars ($\S$ 5), estimate the
distance to the cluster via several methods based on different families of variable
stars ($\S$ 6), discuss the structure of the Horizontal Branch ($\S$ 7), and
summarize our results ($\S$ 8). Finally, in Appendix A we discuss the properties and
classification of a number of variables that require further analysis to
characterize them.

\section{Observations and reductions}
\label{sec:ObserRed}

\begin{table}[t]
\footnotesize
\caption{The distribution of observations of NGC~6934.
Columns $N_{V}$ and $N_{I}$ give the number of images taken with the $V$ and $I$
filters respectively. Columns $\MakeLowercase{t}_{V}$ and $\MakeLowercase{t}_{I}$
provide the exposure time,
or range of exposure times. In the last column the average seeing is listed.}
\centering
\begin{tabular}{lccccc}
\hline
Date  &  $N_{V}$ & $t_{V}$ (s) & $N_{I}$ &$t_{I}$ (s)&Avg seeing (") \\
\hline
 2011-08-05 & 16 & 120-140 & 16 & 20-35   & 1.7\\
 2011-08-06 & 20 & 110-170 & 22 & 20-45  & 1.6\\
 2011-08-07 &  6 & 100-200 &  5 & 25-60  & 1.9\\
 2012-10-20 & 20 & 100-200 & 20 & 20-80 & 3.1\\
 2012-10-21 & 40 & 60-80 & 43 & 15-70 & 2.0\\
 2014-08-03 &  4 & 90 &  3 & 30 & 1.7\\
 2014-08-05 & 20 & 70 & 20 & 30 & 1.7\\
 2016-10-02 & 34 & 30  & 36 & 10 & 2.0\\
 2016-10-03 & 38 & 30 & 36 & 10 & 2.1\\
\hline
Total:   & 198&    &  201  &    &\\
\hline
\end{tabular}
\label{tab:observations}
\end{table}

\begin{figure*}[!t]
\begin{center}
\includegraphics[scale=1.3]{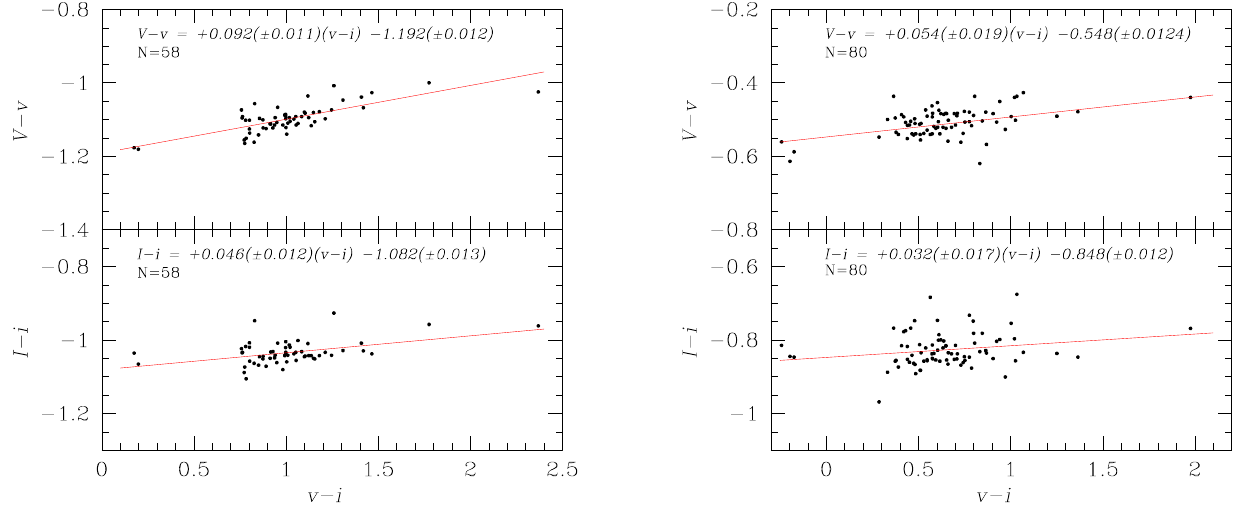}
\caption{The transformation relationship between the instrumental and standard
photometric systems using a set of standards of Stetson (2000) in the FoV of our
images
of NGC~6934 for settings A (left panel) and B (right panel).}
    \label{trans}
\end{center}
\end{figure*}

\subsection{Observations}

The Johnson-Kron-Cousins $V$ and $I$ observations
used in the present work were obtained between  August 2011 and October 2016 with
the 2.0m-telescope at the Indian Astronomical Observatory (IAO), Hanle,
India, located at 4500~m above sea level in the Himalaya.
The detector used in HFOSC is 4K x 2K CCD
    using a SITe002 chip with pixel size of 15$\mu ^{2}$ with imaging area
    limited to 2048 x 2048 pixels and an image scale of 0.296 arsec/pixel, translating
to a field of view (FoV) of approximately
10.1$\times$10.1~arcmin$^2$. Our data consist of 198~$V$ and 201~$I$ images. Table
\ref{tab:observations}
gives an overall summary of our observations and the seeing conditions.

\subsection{Difference Image Analysis}
\label{DIA}

Image data were calibrated using bias and flat-field
correction procedures. We used the Difference Image Analysis (DIA)
to extract high-precision time-series photometry in the FoV of NGC~6934. We used
the
{\tt DanDIA}\footnote{{\tt DanDIA} is built from the DanIDL library of IDL routines
available at \texttt{http://www.danidl.co.uk}}
pipeline for the data reduction process (Bramich et al.\ 2013), which includes an
algorithm that models the convolution kernel matching the PSF
of a pair of images of the same field as a discrete pixel array (Bramich 2008).
A detailed description of the procedure is available in the paper by
 Bramich et al.\ (2011), to which the interested reader is referred for
the relevant details.

We also used the methodology developed by
Bramich \& Freudling (2012) to solve for the
magnitude offset that may be introduced into the photometry
by the error in the fitted value of the photometric scale factor
corresponding to each image.
The magnitude offset due to this error was small, of the order
of $\approx 2{-}5$~mmag.

\subsection{Transformation to the \textit{VI} standard system}

While in the season of 2011 the cluster was centered in the CCD, in 2012, 2014 and
2016
we observed the cluster slightly off center in the CCD in order to avoid
the very bright star to the west of the cluster. We shall refer to these two
setting as A and B.
We have treated these setting as independent, each with its own set of standard
stars and transformation equations to the standard system.

From the standard stars of Stetson (2000)\footnote{%
 \texttt{http://www3.cadc-ccda.hia-iha.nrc-cnrc.gc.ca/\\
community/STETSON/standards}}
in the field of NGC~6934, we identified 58 and 80 standard star in the FoV of our
settings A and B respectively, with $V$ in the range 13.9--20.6~mag and $V-I$
within $-0.17$--$2.37$~mag. These stars were used to transform our instrumental system
to the Johnson-Kron-Cousins photometric system (Landolt 1992). The standard minus the
instrumental magnitude differences show a mild dependence on the colour as displayed
in Fig.~\ref{trans} for both settings. The transformation equations are of the form:
 %
%\begin{equation}
%V= v +0.0440(\pm0.0076)(v-i) -1.2353(\pm0.0054),
%\label{eq:transV}
%\end{equation}
%
%\begin{equation}
%I= i +0.0190(\pm0.0089)(v-i) -1.7821(\pm0.0064).
%\label{eq:transI}
%\end{equation}

\noindent
For setting A;

\begin{eqnarray}\label{eq:transV}
V &=& v +0.092\, (\pm0.011)\, (v-i) \nonumber \\
  && -1.192\, (\pm0.012),
\end{eqnarray}
\begin{eqnarray}\label{eq:transI}
I &=& i +0.046\, (\pm0.012)\, (v-i) \nonumber \\
  && -1.082\, (\pm0.013).
\end{eqnarray}

\noindent
For setting B;, in 2012, 2014 and 2016 we
observed the cluster slightly off center in the CCD.

\begin{eqnarray}\label{eq:transV}
V &=& v +0.054\, (\pm0.019)\, (v-i) \nonumber \\
  && -0.548\, (\pm0.012),
\end{eqnarray}
\begin{eqnarray}\label{eq:transI}
I &=& i +0.032\, (\pm0.017)\, (v-i) \nonumber \\
  && -0.848\, (\pm0.012).
\end{eqnarray}

We note the zero point differences in the above transformation equations for each of
the two settings, for both filters. This is in spite having used a large number of
standard stars in common. The reason for this is that, for each setting a different
reference image was used, each with ist own quality. The zero point offsets indicate
that one of the reference images was built from images taken under different
transparency conditions, thus producing two
independent instrumental magnitude systems. Once the instrumental magnitudes of
each setting are converted into the standard system, the light curve matching between
the two settings is very good, as can be seen in Figs. \ref{RRL_A} and \ref{RRL_C}.

\section{Variable Stars in NGC 6934}

The variable stars in our FoV are listed in Table~\ref {variables}
along with their
mean magnitudes, amplitudes, and periods derived from our photometry. The coordinates
listed in columns~10 and~11 were taken from the CVSGC, and were calculated by KOS01.
For comparison we include in column 7, the periods as listed by KOS01, and it should
be noted that in some cases the periods are significantly different from the ones
found from our data. For stars with light curves poorly covered by our photometry we
have adopted the period of KOS01. The light curves of the RR Lyrae stars are shown in
Figs. \ref{RRL_A} and \ref{RRL_C}. Unfortunately, due to poor seeing conditions in
some nights of setting B, we were not able to calculate an accurate reference flux,
hence we could not recover the magnitudes scale light curve for some variables
with particularly bad blending conditions.

\begin{figure*}
\includegraphics[scale=0.95]{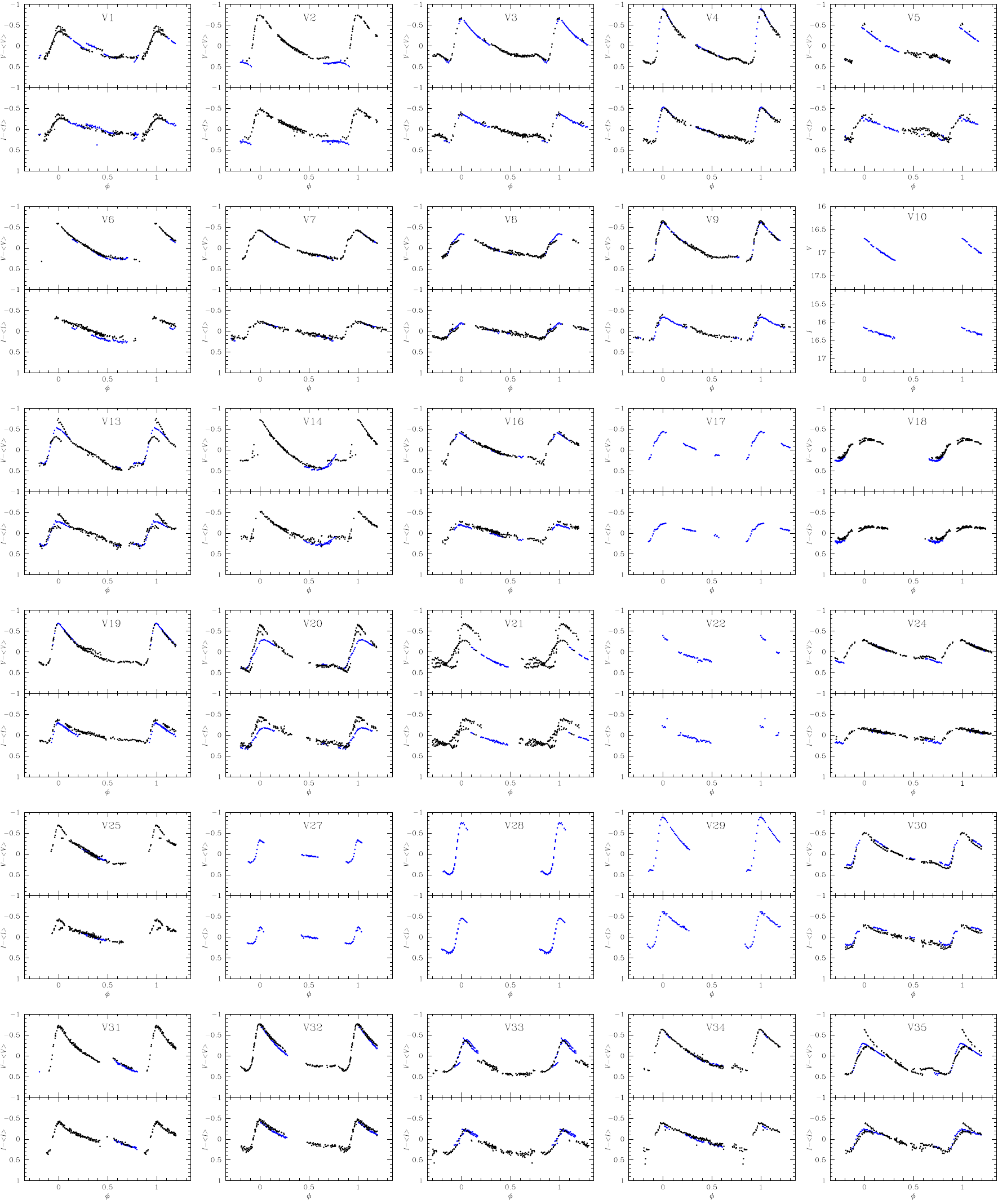}
\caption{Light curves of the RRab stars in our FoV
phased with the periods listed in Table~\ref{variables}. In order to appreciate the
amplitude differences from star to star, the vertical axis displays $V-<V>$ and
$I-<I>$. The intensity-weighted means are listed in Table \ref{variables}. Blue
symbols correspond to data from setting A and black to data from setting B. See text
in the Appendix A for discussion of some individual stars.}
    \label{RRL_A}
\end{figure*}

\begin{figure*}
%\ContinuedFloat
\includegraphics[scale=0.95]{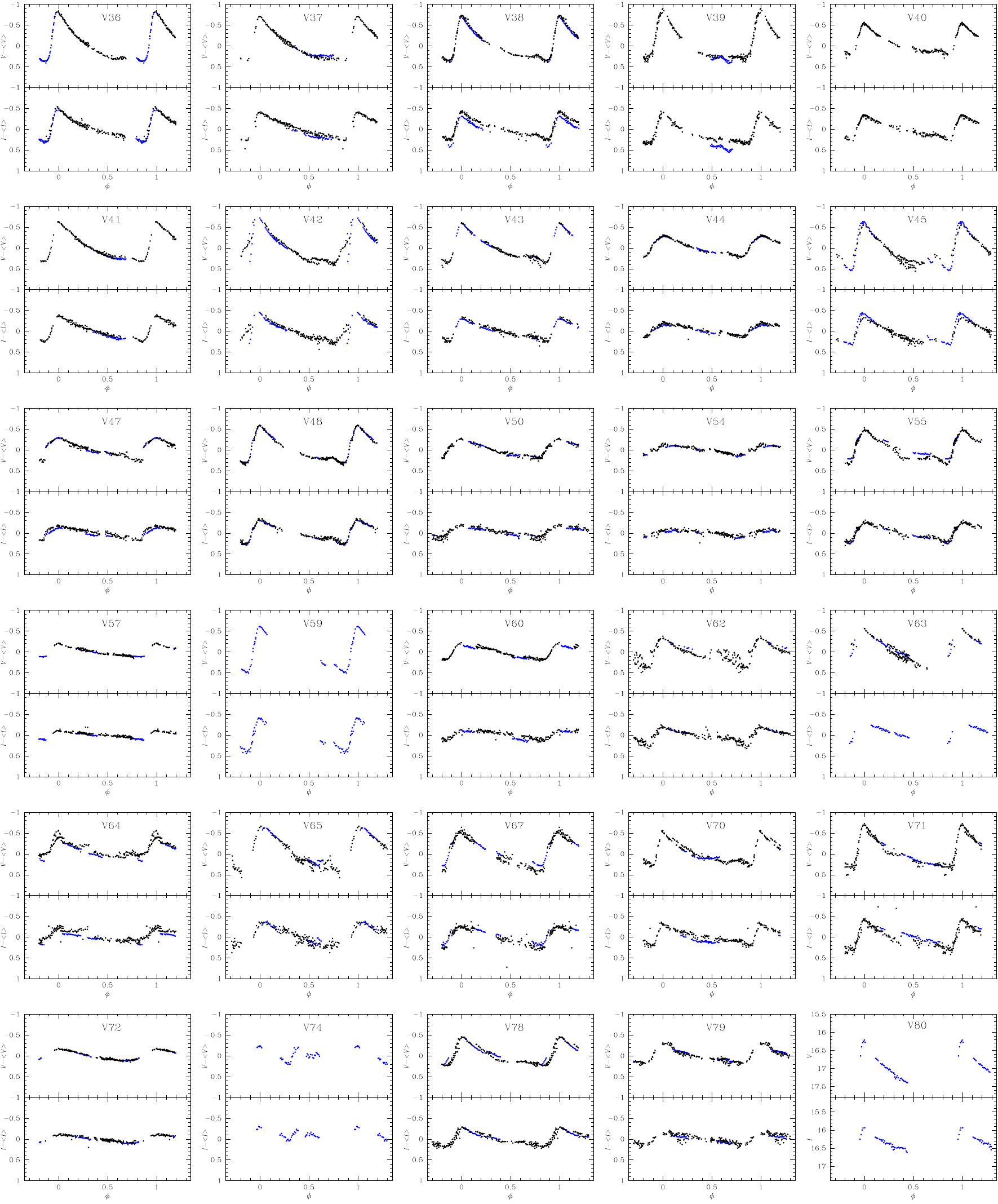}
\caption{Continued}
\end{figure*}

\begin{figure*}
%\ContinuedFloat
\includegraphics[scale=0.95]{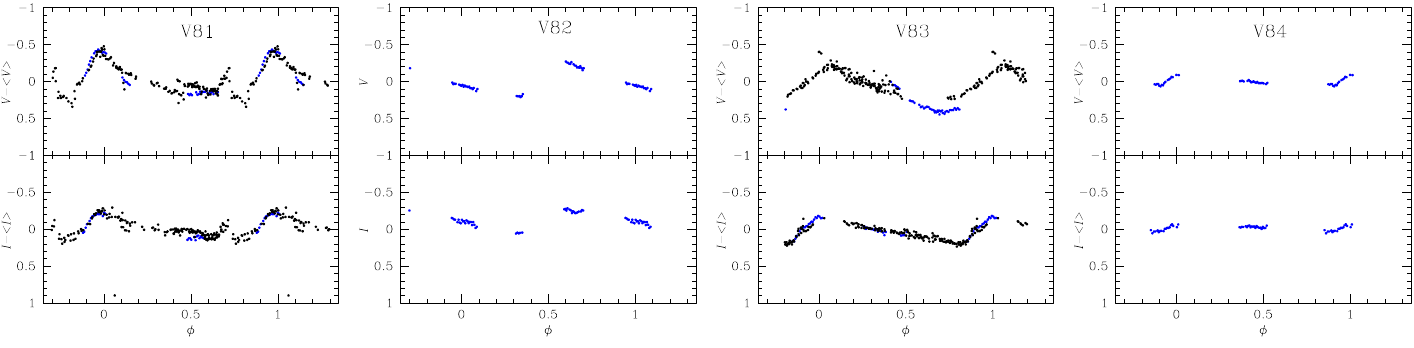}
\caption{Continued}
\end{figure*}

\begin{figure*}
\includegraphics[scale=0.95]{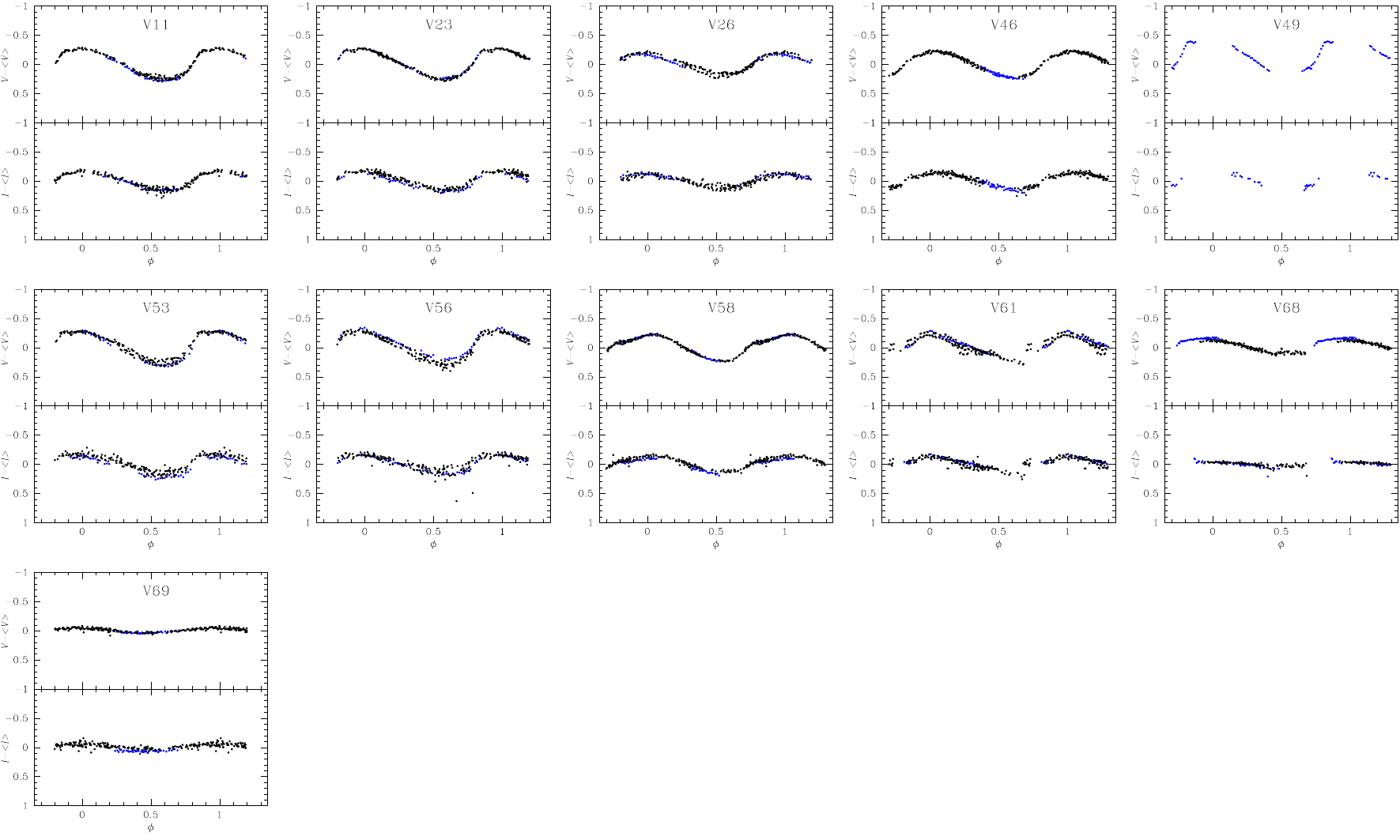}
\caption{Same as Fig \ref{RRL_A} for RRc stars.}
    \label{RRL_C}
\end{figure*}

% TABLA 1
\begin{table*}
\scriptsize
\begin{center}

\caption{General data for all of the confirmed variables in NGC~6934 in the \fov of
our images.  Previous period estimates for each variable from KOS01 are listed in
column~7 for comparison with our periods reported in column~9.}
\label{variables}

\begin{tabular}{llllllllllll}
\hline
Variable & Variable & $<V>$ & $<I>$   & $A_V$  & $A_I$   & $P$ (KOS01) &
HJD$_{\mathrm{max}}$
& $P$ (this work)    & RA   & Dec         \\
Star ID  & Type     & (mag) & (mag)   & (mag)  & (mag)   & (d)  &
($+2\,450\,000$) & (d) & (J2000.0)   & (J2000.0)    \\
&&&&&&&&&&&\\
\hline
V1      & RRab Bl    & 16.80  &16.28  & 0.805       &0.638  & 0.56751
       & 6875.4072      & 0.600391     & 20:34:08.4 & 07:23:39 \\
V2       & RRab     & 16.952  & 16.398  &  1.093  & 0.726   &0.481947
        & 7665.0944    & 0.481992      & 20:34:08.6 &  07:24:02\\
V3       & RRab     & 16.912  & 16.320  &  1.072   & 0.715   &0.539806
        & 7664.2940   & 0.539806      & 20:34:11.4 &  07:25:15\\
V4       & RRab     & 16.775  & 16.167  &  1.325   & 0.862   &0.616422
        & 5780.3908    & 0.616415      & 20:34:13.9 &  07:25:15\\
V5       & RRab      & 16.90 &16.27 &0.910     & 0.600  & 0.564560
        & 7664.2520       &  0.559514     & 20:34:15.2  & 07:27:58\\
V6       & RRab      & 16.985 &16.289  &0.932     & 0.611  & 0.555866
        & 6222.0656       & 0.555847      & 20:34:09.5 &  07:23:45\\
V7       & RRab      & 16.872 & 16.212 &0.700     & 0.439  & 0.644049
        & 6221.1532     & 0.644047   & 20:34:17.4 &  07:25:16\\
V8       & RRab Bl    &16.893   & 16.222  & 0.576 & 0.385   & 0.623984
       & 5780.4479     & 0.620522   & 20:34:18.0 &  07:25:08\\
V9      & RRab      &16.965   & 16.320  &0.967   & 0.6338  & 0.549156
       & 6875.4291       & 0.549157     & 20:34:15.6  & 07:24:36\\
V10      & RRab      &--   & --  &--   & --  & 0.519959
       & 5781.3595      & 0.518391    & 20:34:02.2  & 07:25:28\\
V11       & RRc     & 16.912 &16.434 &  0.553  & 0.345   & 0.30867
        & 7664.1126    &0.308679   & 20:34:12.5  & 07:24:46\\
V12       & RRab     &16.5  & --  & 0.978    & --  & 0.464215
        & 7665.2551     &0.894809   & 20:34:13.2  & 07:23:34\\
V13       & RRab Bl    & 16.87  &16.26  & 1.227  & 0.778   & 0.551334
       & 6222.1033      & 0.551333  & 20:34:08.1  & 07:24:42\\
V14       & RRab      & 16.80  &16.309 &1.162  & 0.785  & 0.521990
       & 6222.0980   & 0.519406  & 20:34:10.9 &  07:22:47\\
V15       & SR?      & 13.773  &12.068 &--  &  --  & --
       & --     & --     & 20:34:11.9 & 07:23:25\\
V16       & RRab      & 16.896  &16.264 &0.768  & 0.462  & 0.604853
       & 6873.3918    & 0.604866    & 20:34:13.7 &  07:24:36\\
V17      & RRab    & 16.927 &16.270 & 0.766  & 0.443   &  0.598272
      & 5780.4430       & --   & 20:34:06.5  & 07:22:30\\
V18       & CWB    & 16.551  & 15.844 & 0.547    & 0.358   & 0.956070
       & 6875.3521      & --   & 20:34:14.6 &  07:24:09\\
V19       & RRab    & 16.645  &15.95 & 1.014    & 0.557   & 0.480550
       & 5780.3621       & 0.480569    & 20:34:13.2 &  07:24:19\\
V20       & RRab Bl   &16.85  & 16.30  & 1.128    & 0.764   & 0.54833
       & 7665.2570       & 0.548224     & 20:34:09.6 & 07:24:34\\
V21       & RRab Bl    & 16.95 & 16.31  &  1.051   & 0.650   &0.526829
        & 6873.4211     & --    & 20:34:08.9  & 07:24:15\\
V22       & RRab      & 16.93 & 16.30  & --    & --   &  0.574280
        & 5781.3595       & 0.545104    & 20:33:55.3 &  07:21:24\\
V23       & RRc     &16.878  &16.409 & 0.560  & 0.368   & 0.28643
       & 6221.1657     & 0.286431   & 20:34:09.3 & 07:24:00\\
V24       & RRab     &16.949  &16.254 & 0.492  & 0.312   & 0.641670
       & 7665.1914    & 0.641673  & 20:34:13.8 &  07:23:24\\
V25       & RRab    &16.893 & 16.31   &1.000   & 0.741  & 0.509086
       & 7664.0792    & 0.509014     & 20:34:14.7  & 07:24:54\\
V26       & RRc    &16.943   & 16.547  &0.431   & 0.273  & 0.259318
       & 7664.0931    & 0.259318     & 20:34:13.4  & 07:21:02\\
V27       & RRab    &16.994   & 16.30   &0.549   & 0.403  & 0.592204
       & 5781.3595    & 0.637029       &  20:34:01.4  & 07:27:39\\
V28       & RRab   &16.80   & 16.28  &  1.260  & 0.844  & 0.485151
        & 5780.4334     & 0.485202  & 20:33:55.6 &  07:25:56\\
V29       & RRab   &17.04   & 16.56  &  1.297  & 0.890  & 0.454818
        & 5779.4387    & 0.455798   & 20:34:05.7  & 07:21:14\\
V30       & RRab   &16.916     &16.265  &  0.871  & 0.585  & 0.589853
        & 7665.0688     & 0.589853   & 20:34:22.1  & 07:26:26\\
V31       & RRab   &16.920    &16.320  &  1.172  & 0.749  & 0.505070
        &  7665.0944    & 0.505780   & 20:34:21.1  & 07:22:37\\
V32       & RRab   &16.942   & 16.349  &  1.135  & 0.733  & 0.511948
        & 6222.1420    & 0.511950   & 20:34:10.6 &  07:25:08\\
V33       & RRab   &16.91    & 16.34  &  0.874  & 0.674  & 0.518445
        & 6221.1771     &  0.507833  & 20:34:13.8 &  07:24:29\\
V34       & RRab   &16.991    & 16.332  &  1.003  & 0.673  & 0.560103
        & 6222.1441    & 0.560099  & 20:34:09.9 &  07:24:30\\
V35       & RRab Bl    &16.75  & 16.25  & 1.104   & 0.707  &0.544222
        & 7665.0587     &0.544220   & 20:34:21.9 &  07:21:56\\
V36       & RRab     & 16.884   & 16.301  &1.181  & 0.833  &0.495659
       & 6875.3708      & 0.495660   & 20:34:12.1  & 07:23:41\\
V37       & RRab      &17.009  &16.336 & 1.056  & 0.671 &  0.533186
       & 6222.1268     &0.533188     & 20:34:12.9 &  07:24:28\\
V38      & RRab    &16.907 &16.266 &1.126   & 0.707   & 0.523562
       & 7665.1914    &0.523559 &20:34:12.2  & 07:23:59\\
V39       & RRab     &16.983 &16.26& 1.212  & 0.777  & 0.502578
       & 7665.2055       &0.504174   & 20:34:11.9 &  07:24:00\\
V40       & RRab     &16.616 & 16.166  &  0.813   & 0.616   &0.560755
        & 7664.0931    & 0.560781     &20:34:10.7  & 07:24:44\\
V41       & RRab     & 16.980 &16.348  &  0.957   & 0.649   &0.520404
        & 7664.1458    & 0.520446    &20:34:13.3  & 07:23:38\\
V42       & RRab     & 16.869 &16.304  &  1.149   & 0.790   &0.524235
        & 5780.3575    & 0.528068    &20:34:15.0 &  07:24:39\\
V43       & RRab     & 16.969 & 16.285  &  0.965   & 0.573   &0.563218
        & 7664.2337    & 0.563183     &20:34:12.8 &  07:24:45\\
V44       & RRab      &16.941 &16.280 &  0.505    & 0.354 & 0.630384
        &6875.3908       &0.630383    &20:34:08.5 & 07:23:48\\
V45       & RRab     &16.80   & 16.30  & 1.178 & 0.761   & 0.53660
       & 5779.3870     & 0.540324   &20:34:09.2  & 07:24:08\\
V46       & RRc     &16.933   &16.455  & 0.441 & 0.303   &0.328557
       & 6222.1441     & 0.328557   & 20:34:12.3  & 07:23:53\\
V47       & RRab     &16.921   & 16.229  & 0.422 & 0.316   & 0.640938
       & 6222.0580   & 0.620252   &20:34:12.0  & 07:23:52\\
V48       & RRab     &16.922  &16.275 & 0.895 & 0.626   & 0.561299
       & 6222.2406     & 0.561319&20:34:13.5 &  07:25:08\\
V49       & RRc      & 16.98   &16.5   &--   & --  &0.285460
       &6222.1215     &0.399840   & 20:34:12.2  & 07:23:22\\
V50       & RRab      &16.984   & 16.32  &  0.507&0.368  &0.634510
        & 7664.1753    & 0.614237   & 20:34:12.4  & 07:23:41\\
V51       & RRab     &--   & --  & -- & --   & 0.564769
       & 6875.4456   & 0.516442   &20:34:11.8 &  07:24:52\\
V52       & SX Phe     &18.943   & 18.482  &0.471 & 0.330  & 0.05976
       &6875.4396    & 0.063563 &20:34:18.3  & 07:22:14\\
V53       & RRc      &16.973   &16.489   &0.600   &0.333 &0.28235
       &7665.1038    &0.282377   & 20:34:13.6 &  07:24:00\\
V54       & RRab      &16.750   & 16.051  &  0.282& 0.210 &0.59020
        & 6873.4027    & 0.764917   & 20:34:12.6 &  07:24:35\\
V55       & RRab     &16.998  & 16.276 & 0.824  & 0.508  & 0.77828
        & 7664.1439      &0.590251   & 20:34:13.3 &  07:24:27\\
V56       & RRc Bl    & 16.972 & 16.467 &0.597   & 0.322  & 0.29104
       & 5780.3667     &0.291054  & 20:34:12.5 &  07:24:18\\
V57       & CWB?      & 15.839  &15.000  & 0.331  & 0.224  & 0.68712
       & 7664.0931     &0.687174   &20:34:12.4 &  07:24:10\\
V58      & RRc    &16.757  & 16.205 & 0.452 & 0.264   & 0.40082
       & 6221.2010   & 0.398628   & 20:34:12.1 &  07:25:05\\
V59       & RRab    &16.928  & 16.441 & 1.125  & 0.831  & 0.53855
        & 5779.4104      &0.538044   & 20:34:12.0  & 07:24:15\\
V60       & RRab     &16.895 & 16.22 &0.397   & 0.25 & 0.66040
       & 7664.29398     &0.654264  & 20:34:12.0 &  07:24:56\\
\hline
\end{tabular}
\end{center}
\end{table*}

% TABLA 1
\begin{table*}
%\ContinuedFloat
\scriptsize
\begin{center}

\caption{Continued}
\label{variables}

\begin{tabular}{llllllllllll}
\hline
Variable & Variable & $<V>$ & $<I>$   & $A_V$  & $A_I$   & $P$ (KOS01) &
HJD$_{\mathrm{max}}$
& $P$ (this work)    & RA   & Dec         \\
Star ID  & Type     & (mag) & (mag)   & (mag)  & (mag)   & (d)  &
($+2\,450\,000$) & (d) & (J2000.0)   & (J2000.0)    \\
&&&&&&&&&& & \\
\hline
V61       & RRc     & 16.890  &16.305  & 0.544  & 0.284  & 0.528
       & 5780.3574     &0.355146   &20:34:11.9  & 07:23:10\\
V62      & RRab    & 16.167 & 15.755 & 0.715 & 0.452 & 0.53067
       & 6873.4210    &  0.530633   & 20:34:11.7 &  07:24:26\\
V63      & RRab   &16.942  & 16.23 & 1.095 & 0.45 & 0.57564
        & 7665.0587     &0.562610   & 20:34:11.6  & 07:24:26\\
V64      & RRab  Bl    & 16.555 & 15.90 &0.641   & 0.433 & 0.57102
       & 7664.2847     &0.567969 & 20:34:11.4  & 07:24:18\\
V65      & RRab     & 16.988   &16.346 & 1.156  & 0.606  & 0.65905
       & 6221.2514    &0.640858   &20:34:11.4 &  07:24:15\\
V66      & RRab?    & --  & -- & -- & --   & 0.54078
       & 5781.3805    & --  & 20:34:11.1  & 07:24:16\\
V67       & RR?     & 17.223   &16.468  & 0.935  & 0.525  & 0.61333
       & 6222.1595     &0.613381   &20:34:10.9 &  07:23:55\\
V68      & RRc    & 16.432  &15.48 & 0.520 & --   & 0.33534
       & 7665.0778    & 0.335344   & 20:34:10.9 &  07:24:00\\
V69       & RRc ?   &16.925  &16.502 & 0.41 & 0.12 & 0.24700
        & 6221.2010    &0.245633 & 20:34:10.8  & 07:23:15\\
V70       & RRab    & 16.773 & 16.088 &0.845   & 0.557 & 0.53935
       & 7665.1291     &0.528880 & 20:34:10.7 &  07:24:06\\
V71       & RRab     & 16.953   &16.373  & 1.041  & 0.769  & 0.57269
       & 6222.1595   &0.563084   &20:34:10.7  & 07:23:54\\
V72      & RRab    & 16.885  & 16.208 & 0.277 &0.206  & 0.66785
       & 7664.0914    & 0.672084  & 20:34:10.5 &  07:25:20\\
V73       & RRd$^{a}$     & 17.01   &16.43 & 0.431  &0.284  & 0.50621
       &  7665.2983    &--   &20:34:09.8  & 07:24:47\\
V74      & RRab    & 17.0  & 16.5 & 0.46&0.33   & 0.56813
       & 6873.4210    & --   & 20:34:09.3 &  07:24:08\\
V75       & EW   &16.91   & 16.17 & 0.27 & 0.23  & 0.28207
        &5780.4004   &--   & 20:34:02.8  & 07:19:35\\
V76       & EW    & 17.94  &  17.62 &0.34 &0.22  &  0.33649
       & 5779.4387   &-- & 20:33:54.6 &  07:19:50\\
V77       & L     & --   &--  & --  & --  & --
       & --    &--  &20:34:10.5 &  07:24:24\\
V78      & RRab    & 16.492  & 16.022 & 0.694 & 0.484  &  0.54230
       & 6222.1846    & 0.557881  & 20:34:12.1 &  07:24:38\\
V79       & RRab     & 16.843  &16.130 & 0.503  & 0.298  & 0.62187
       &7664.2355     &0.638841   &20:34:10.5  & 07:24:24\\
V80      & RRab    & --  & -- & -- & --   & 0.54427
       & 5781.3805   &  0.542778 & 20:34:11.7 &  07:24:07\\
V81       & RRab   &16.92  & 16.19 &0.69 &0.40  & 0.57262
        & 6875.4456   &0.617816 & 20:34:11.2  & 07:24:23\\
V82       & RRab    & 16.9& 16.3 &--   & -- & 0.73113
       & 7665.2486     &0.73113 & 20:34:10.7 &  07:24:17\\
V83       & RRab     & 16.70   &16.20 & 0.662  & 0.399  & 0.54055
       & 6222.0580   &0.529951   &20:34:11.2 &  07:24:26\\
V84      & RRab    &16.535  &15.79 & 0.142 &0.105 & 0.66535
       & 5779.4387    & 0.672084  & 20:34:12.1 &  07:24:28\\
V85       & ?     & 17.316  &16.293  &0.096 &0.081 & 1.622
       & 7664.2817  &1.6429   &20:34:31.0 &  07:21:5\\
V86      & SR    &13.78  &12.08 & -- &--  & $\sim$49.
       & --    & --  & 20 34 19.5 &07 22 50\\
V87$^{a}$      & CWB   & 14.474  & 13.332 & 0.115 &0.066 & --
       & 6221.1684    & 0.574663 & 20:34:12.6 & 07:24:12\\
V88$^{a}$       & RRab     & 16.746  &16.367 &1.095 &0.781  & --
       & 7664.0652   &0.519621   &20:34:11.4 & 07:24:10\\
V89$^{a}$      & RRab   & 17.122  & 16.385 & 0.999 &0.768  & --
         & 5779.3729    &0.525269 & 20:34:11.3 &07:24:11\\
V90$^{a}$     &CWB &17.005   & 16.302 & 0.090 &0.05 &--
       &7664.0914    &1.056153  & 20:34:12.4 &07:20:53\\
V91$^{a}$      &RRab &16.90   & 16.35 & 0.912 &0.453 &--
       &5780.4143   &0.547098 & 20:34:11.8 &07:24:11\\
V92$^{a}$    &SX Phe &19.403   & 18.96 & 0.12 &0.09 &--
       &5779.3963  &0.045858 & 20:34:05.0 &07:25:27\\
V93$^{a}$      &SX Phe &18.638 &-- &0.141&--&--
       &5779.42926   &0.099016& 20:34:10.1 &07:24:11\\
V94$^{a}$    &RRab &15.80   &15.28 & 0.43 &0.30 &--
       &5781.3595  &0.573329&20:34:11.8  &07:24:16\\
V95$^{a}$    &SX Phe &19.827   &19.308  & --&-- &--
       &5780.4631  &0.0243125 &20:34:00.9 &07:19:58\\
V96$^{a}$      &SR & 13.814 & 12.694 &0.18&0.1&--
       &7664.0774   &9.54& 20:34:22.2 &07:20:10\\
V97$^{a}$    &SR &14.079   &12.948 & 0.07 &-- &--
       &--  &--&20:34:16.9&07:20:45 \\
V98$^{a}$    &SR & 13.782 &12.566 &0.06&-- &--
       &-- &-- &20:34:23.9&07:27:40\\

\hline
C1$^{a}$      &RRab? &17.76   &16.65 & 0.20 & 0.35:&--
       &6222.1595  &0.523307 & 20:34:42.3 &07:28:29\\
C2$^{a}$      &SX Phe &19.96   &19.51 & 0.09 & 0.18&--
       &5780.4479&0.06961 &20:34:13.3 &07:23:32\\
C3$^{a}$      &SX Phe &19.968   &19.6 & 0.18 & 0.15:&--
       &5779.4293   &0.061280 & 20:34:10.1 &07:24:55\\
C4$^{a}$      &SX Phe &19.936   &19.566 & -- & --&--
       &5780.4631&0.03953 &20:34:03.8 &07:24:35\\
              &       &         &      &     &   &
       &         &0.06951 &           &         \\
C5$^{a}$      &SX Phe &20.053   &19.494 & -- & --&--
       &5781.3595   &0.052579 &20:34:15.5&07:27:50\\

\hline
\end{tabular}
\raggedright
\center{\quad $^{a}$Newly found in this work. See Appendix A for a discussion.}
\end{center}
\end{table*}

\subsection{Search for new variables}
\begin{figure}
\includegraphics[scale=0.40]{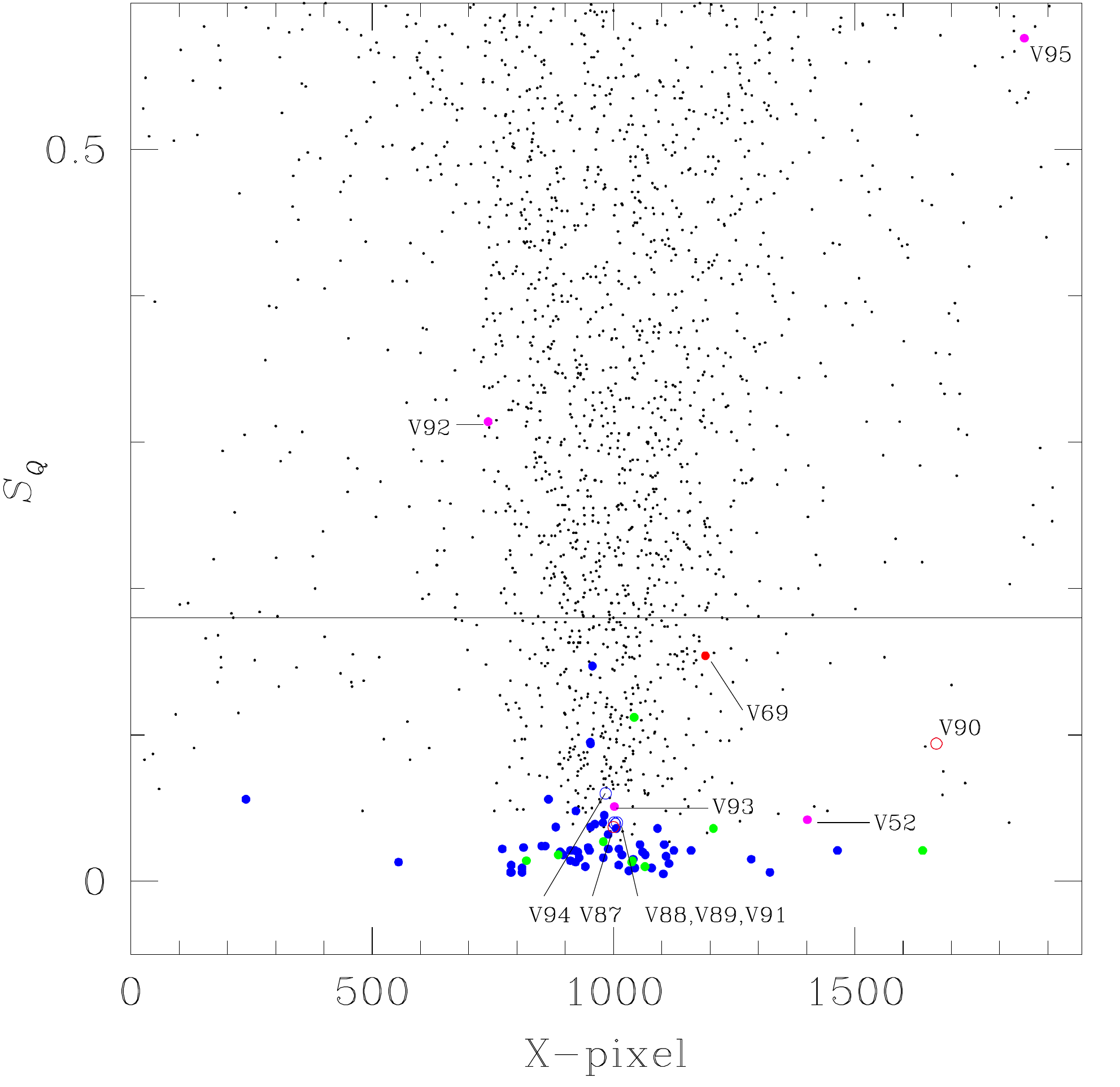}
\caption{Minimum value of the string-length parameter $S_Q$ calculated
for the 2113 stars with a light curve in our V reference image, versus
the CCD X-coordinate. Colours are as in Fig. 2. The horizontal line is an
arbitrarily defined threshold 0.172, below which all known variables are located.}
    \label{SQ}
\end{figure}

We were able to isolate 4274 light curves in $V$ and 4273 in $I$ of individual star
in the FoV of our images for setting A. A search for new variables was conducted using
three different strategies. First we use the string-length method (Burke, Rolland \&
Boy 1970; Dworetsky 1983), in which each light curve was phased
with periods between 0.02 and 1.7 d and a normalized string-length
statistic $S_Q$ is calculated for each trial period. A plot of the minimum values
of $S_Q$ versus the
X-coordinate for each star is shown in Fig. \ref{SQ}. All known variable are
naturally near the bottom of the distribution. We have drawn and arbitrary threshold
at 0.172, below which we find all the known variables. The light curve of
every star below the threshold was explored for variability.

A second approach to the search of new variables consisted in blinking all
residual images. Variable stars are evident by their variation from image to
image. A third approach was to
separate the light curves of stars in a given region of the CMD
where variables are expected, e.g. HB, the blue stragglers region, the upper
instability strip and the tip of RGB. A further detailed inspection of the light
curves of stars in these regions may prove the variability of some stars.

\begin{figure}
\includegraphics[scale=0.38]{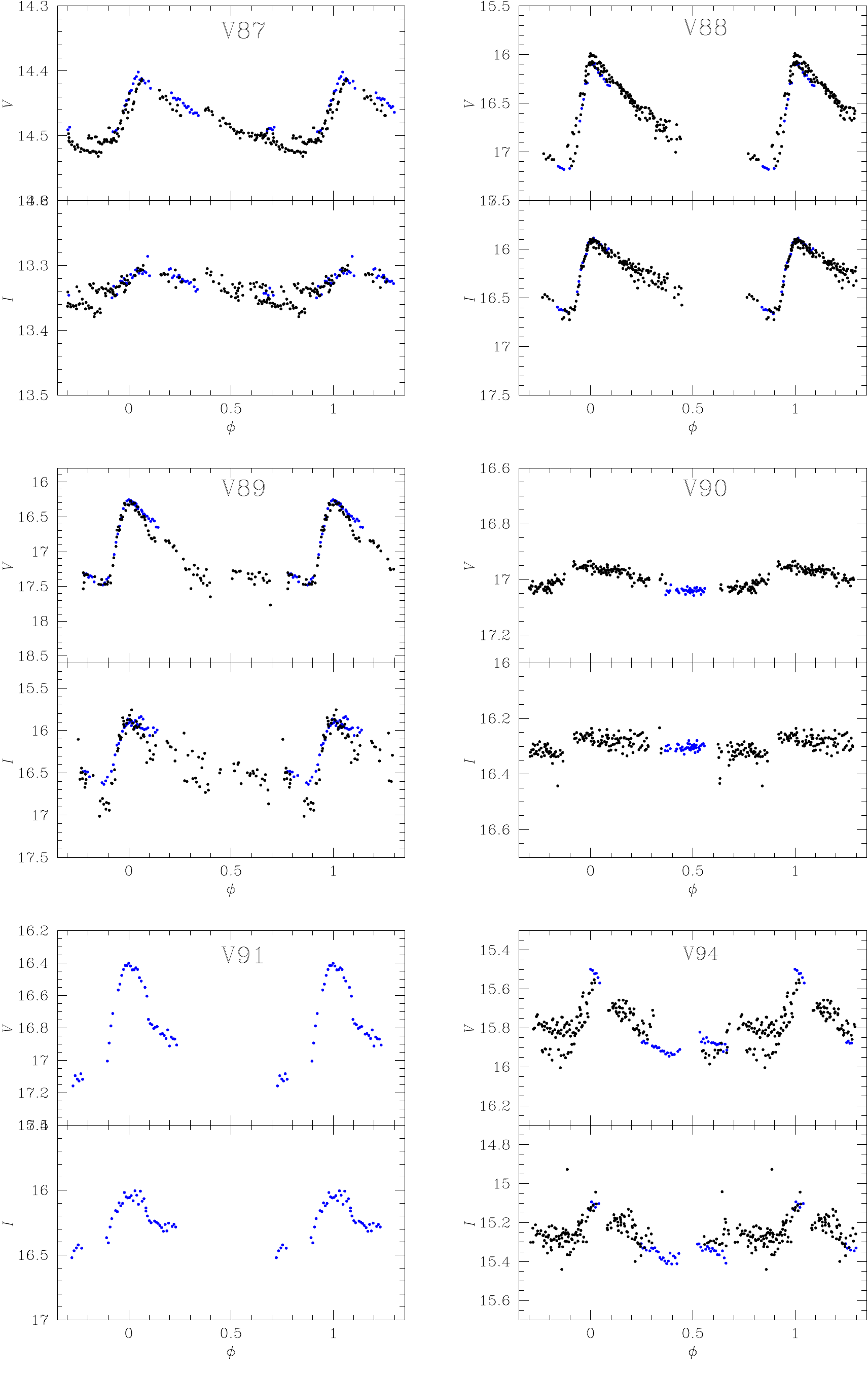}
\caption{Newly discovered RR Lyrae stars and one CW (V90). For an individual
discussion on these stars see the Appendix A.}
    \label{New_var}
\end{figure}

A combination of the three methods described above allowed us to identify all
previously
known variables in the FoV of our images and reveal the existence of twelve new
ones, labeled as V87-V98 and their general properties are listed in Table
\ref{variables}. Four of these are RRab stars (V88, V89, V91, and V94), two
CWB (V87, V90) whose light curves are shown in Fig.\ref{New_var}, three SX Phe
stars (V92, V93 and V95) that shall be further discussed in section $\S$,
\ref{Sec:SXPHE} and three SR or semi-regular variables  (V96, V97
and V98).

We have also identified five star that seem to be variable but given their
position in a crowded region and/or being part of blends their light curves are
dubious and need further confirmation. We have not assigned a variable number to
these candidates but their general properties are however included
in the bottom of Table \ref{variables}, their light curves are shown in Fig.
\ref{Cand} and they are also discussed in Appendix A.

\begin{figure}
\includegraphics[scale=0.40]{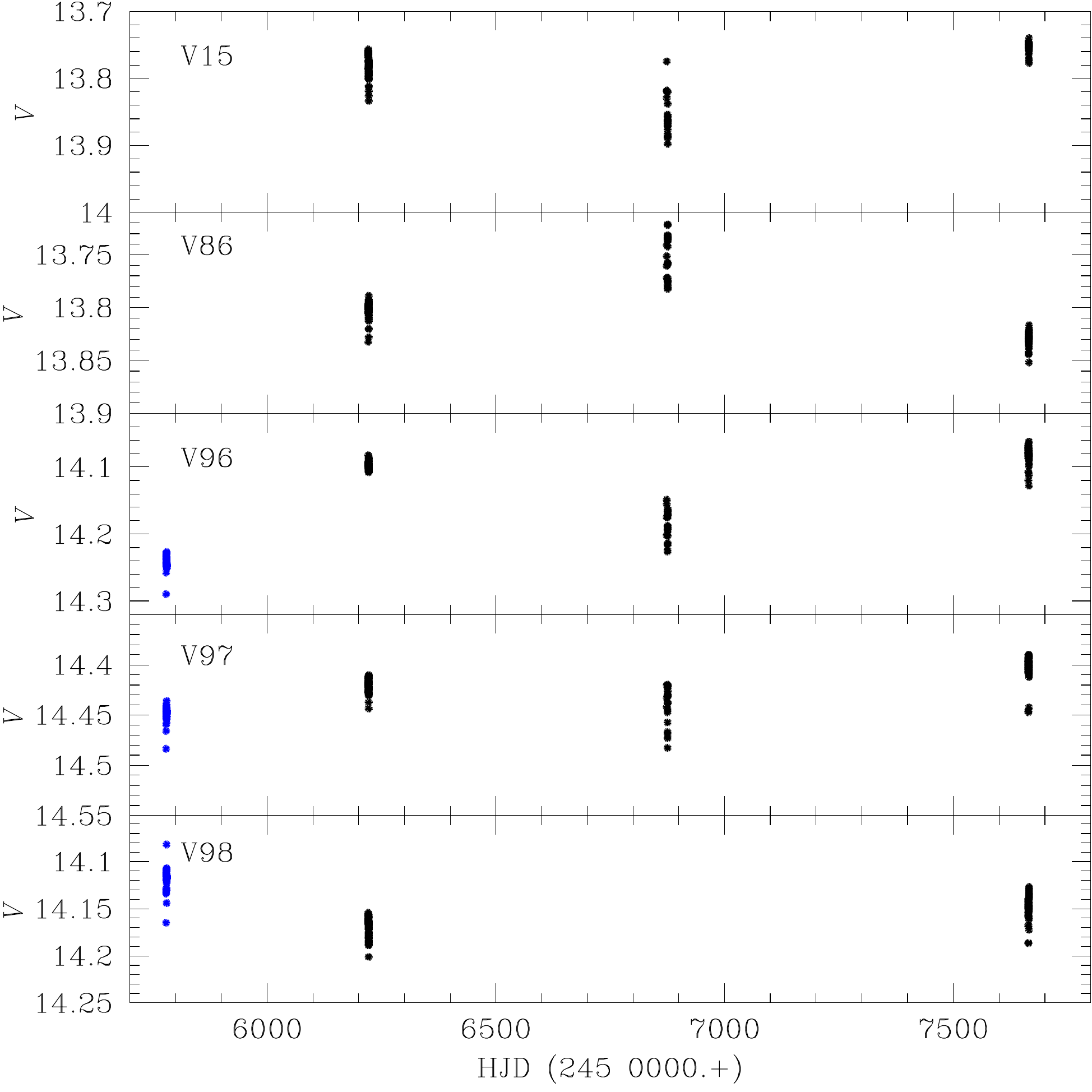}
\caption{Semi-regular (SR) variables near the tip of the Red Giant Branch.
While the variability of V15 and V84 has been reported in the past,
V96, V97 and V98 are new discoveries in this paper.
For an individual discussion on these stars see the Appendix A.}
    \label{SR_HJD}
\end{figure}

\begin{figure}
\includegraphics[scale=0.40]{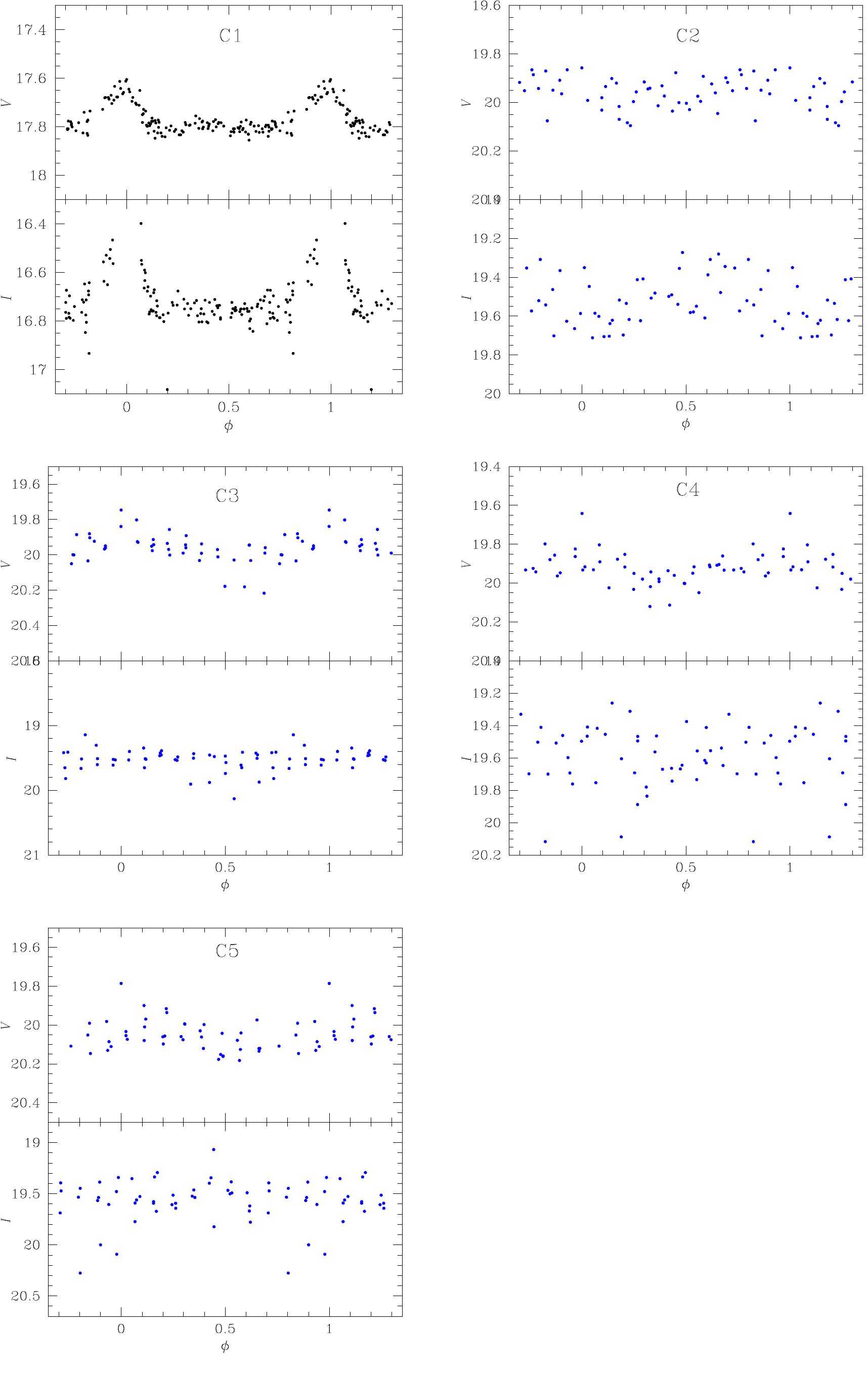}
\caption{Variables whose nature and classification is to be confirmed, phased with the
periods listed in Table \ref{variables}. C4 is likely a double mode SX Phe with
periods 0.03953 d and 0.06951 d.}
    \label{Cand}
\end{figure}

\subsection{The Color-Magnitude diagram}

\begin{figure*}
\begin{center}
\includegraphics[scale=0.7]{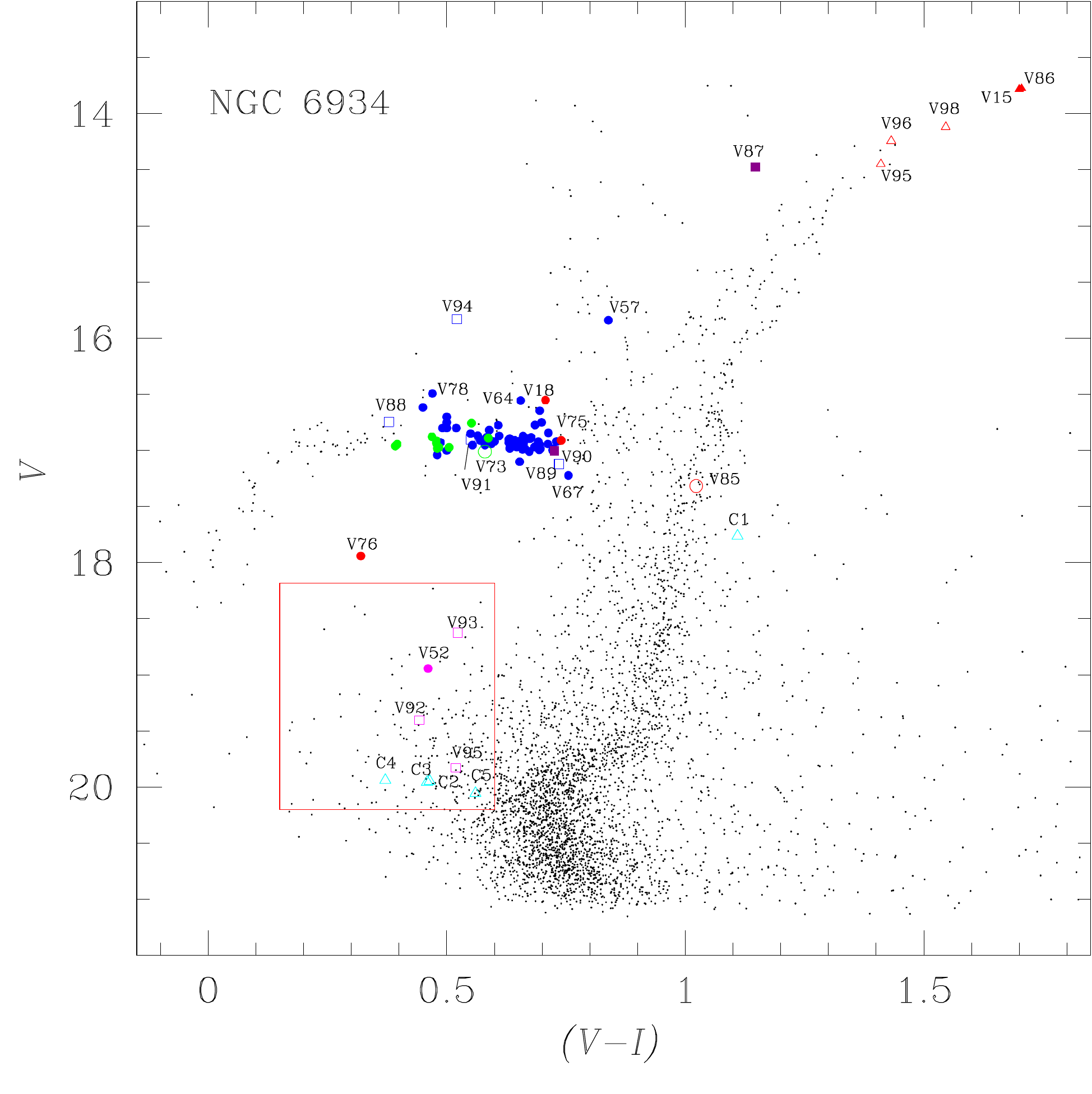}
\caption{CMD of NGC 6934 (setting A). The plotted magnitudes and colors for all
non-variable stars are magnitude-weighted means of their light curves. Variables are
plotted using their intensity-weighted means $<V>$ and $<V>-<I>$. Symbols and
colors are: blue and green circles RRab and RRc stars respectively; cyan
symbols are confirmed SX~Phe stars, turquoise triangles are for candidate SX Phe
stars; red circles EW binaries; purple squares are possible CWB stars (V87, V90);
open green circle is the double mode or RRd star (V73);
red triangles are SR variables. Newly discovered variables are those labeled
from V87 to V95 and plotted with open symbols. The red rectangle is the region where
the search for variability
among the blue stragglers was conducted.}
\label{CMD}
\end{center}
\end{figure*}

\begin{figure*}
\begin{center}
\includegraphics[scale=0.95]{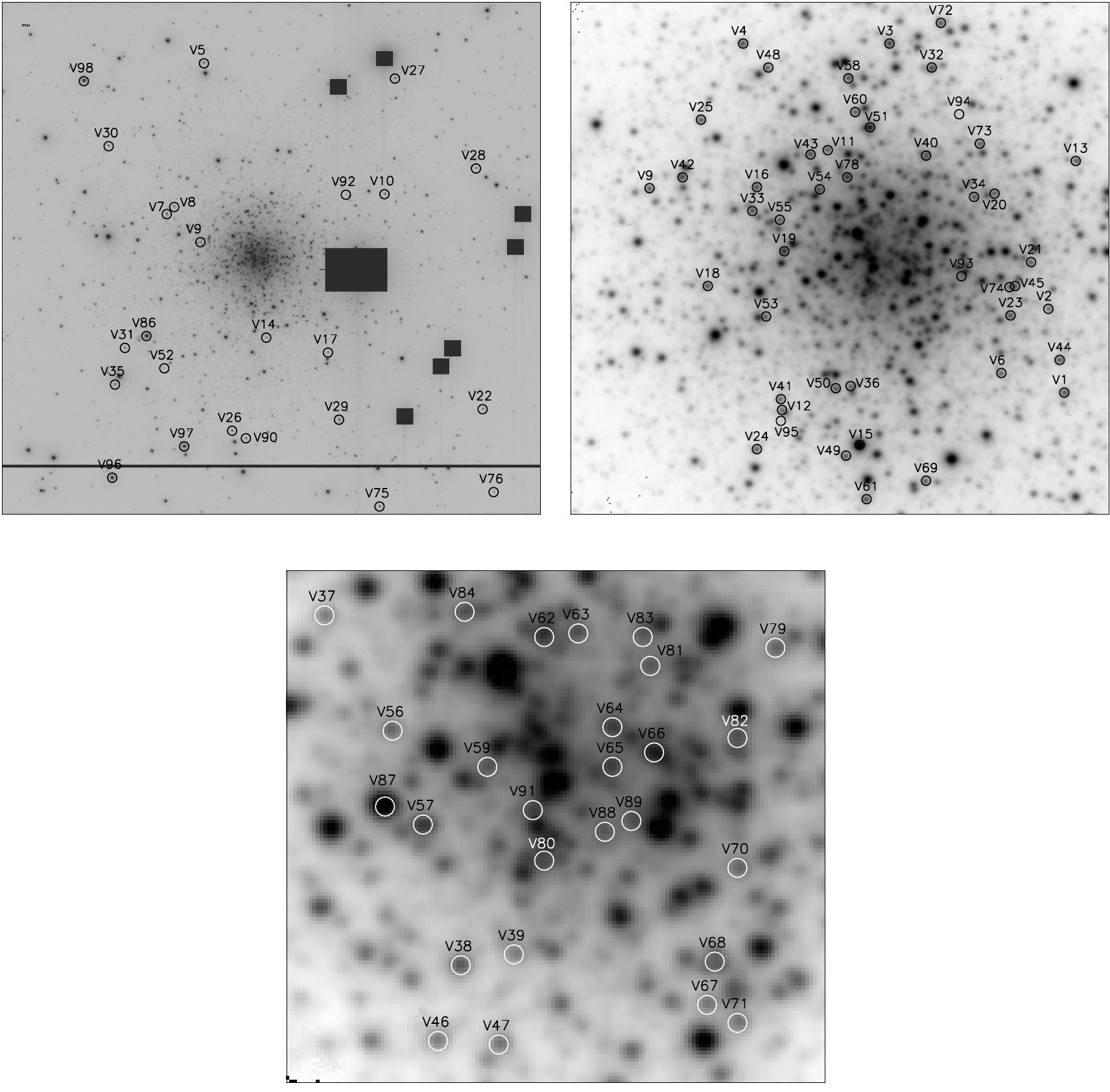}
\caption{Finding chart constructed from 5 of our best $V$ images in setting A.
The top panel
display the complete FoV of our images and it is about $9.6\times9.6$~arcmin$^{2}$.
The middle panel display the central region of the cluster and it is about
$2.3\times2.3$~arcmin$^{2}$. The bottom panel displays the core region of the cluster
and the field is $0.7\times0.7$~arcmin$^{2}$. All stars listed in Table
\ref{variables} are identified in at least one of the panels. In all panels North is
up and East is left.}
\label{chart}
\end{center}
\end{figure*}

The CMD of the cluster is shown in Fig.~\ref{CMD}, where the location of the
known and the newly discovered
variables is marked. All variable stars are plotted using their intensity-weighted
means $< V >$ and corresponding colour $<V> - <I>$.  The finding chart of all
variables in
Table~\ref{variables} is shown in Fig.~\ref{chart} for the peripheral and the
core regions.

%\begin{figure*}
%\includegraphics[width=14.cm,height=14.cm]{SXP_3201.eps}
%\caption{Light curves of the SX~Phe stars in our FoV
%phased with the main period listed in Table~\ref{variables}. The vertical scale
%is not the same for all curves; the tick mark on the vertical axis is equivalent to
%0.1~mag. Variables V122, V123, and V124 are new discoveries.}
%    \label{SXphe}
%\end{figure*}

\section{Physical parameters of RR~Lyrae stars}
\label{sec:Four}

It has been a commonly adopted procedure to calculate [Fe/H], log$(L/{\rm
L_{\odot}})$, and log~$T_{\rm eff}$ for individual RR~Lyrae star by means of the
Fourier decomposition of their light curve, and to employ ad-hoc semi-empirical
calibrations that correlate the Fourier parameters with the physical quantities. The
readers are referred to the work published by Arellano Ferro et al. (2016b; 2017) and
the references therein for a recent example and a summary.

The form of the Fourier representation of a given light curve is:
\begin{equation}
\label{eq.Foufit}
m(t) = A_0 + \sum_{k=1}^{N}{A_k \cos\ ({2\pi \over P}~k~(t-E) + \phi_k) },
\end{equation}
%
%\noindent
where $m(t)$ is the magnitude at time $t$, $P$ is the period, and $E$ is the epoch. A
linear
minimization routine is used to derive the best fit values of the
amplitudes $A_k$ and phases $\phi_k$ of the sinusoidal components.
From the amplitudes and phases of the harmonics in Eq.~\ref{eq.Foufit}, the
Fourier parameters, defined as $\phi_{ij} = j\phi_{i} - i\phi_{j}$, and $R_{ij} = A_{i}/A_{j}$,
are computed.

%\begin{figure}
%\includegraphics[width=8.cm,height=7.cm]{V117raw.eps}
%\caption{$V$ and $I$ light curves of the SR variable V117.
%Intra-night variations are likely due to
%noise from seeing variations. Gradual fainting over the four nights  should be real.}
%    \label{V117}
%\end{figure}

We have argued in previous papers in favour of the calibrations developed by Jurcsik
\& Kov\'acs (1996) and Kov\'acs \& Walker (2001) for the iron abundance and absolute
magnitude of RRab stars, and those of Morgan, Wahl \& Wieckhorts (2007) and Kov\'acs
(1998) for RRc stars.
The effective temperature $T_{\rm eff}$ was estimated using the calibration of Jurcsik
(1998). These calibrations and their zero points have been discussed in detail in
Arellano Ferro et al.\ (2013) and we do not explicitly repeat them here.

The value of $A_0$ and the Fourier light-curve fitting parameters
for 16 RRab and 9 RRc stars with no apparent signs of amplitude modulations
are given in Table~\ref{tab:fourier_coeffs}. These Fourier parameters and the above
mentioned
calibrations were used in turn to calculate the physical parameters listed in
Table~\ref{fisicos}. The absolute magnitude $M_V$ was converted into luminosity with
$\log (L/{\rm L_{\odot}})=-0.4\, (M_V-M^\odot_{\rm bol}+BC$). The bolometric
correction was
calculated using the formula $BC= 0.06\, {\rm [Fe/H]}_{ZW}+0.06$ given by Sandage \&
Cacciari (1990). We adopted  $M^\odot_{\rm bol}=4.75$~mag. For the distance
calculation, we have adopted E(B-V) =0.1 (Harris 1995) given that there are no signs
of differential reddening.

Before the iron calibration of Jurcsik \&
Kov\'acs (1996) for RRab stars can be applied to the light curves, a
``compatibility condition parameter'' $D_m$ should be calculated. {\bf These
authors have made clear that their calibration is only applicable to light curves of
RRab variables that are ‘similar’ to the light curves
that were used to derive calibration. Jurcsik \& Kov\'acs (1996) and  Kov\'acs \&
Kanbur (1998) defined a deviation parameter Dm describing the deviation
of a light curve from the calibrating light curves, based on the Fourier parameter
interrelations.}  These authors advise to
consider only light curves for which $D_m < 3.0$. The values of $D_m$ for each of the
RRab stars are also listed in Table~\ref{tab:fourier_coeffs}. Most of them
fulfill the $D_m$ criterion however, in order to attain a reasonable
size for our sample, we relaxed the criterion and accepted stars with
$D_m < 5.0$ which allowed us to include stars V9 and V48. V50 was hence not included
in the iron abundance calculation.

\begin{table*}
\scriptsize
\begin{center}
\caption[] {Fourier coefficients of  \RRab and \RRc stars in NGC~6934.
The numbers in parentheses indicate
the uncertainty on the last decimal place. Also listed are the number of
harmonics~$N$ used to fit the light curve of each variable and the
deviation
parameter~$D_{\textit{\lowercase{m}}}$.}
\label{tab:fourier_coeffs}
\begin{tabular}{lllllllllrr}
\hline
Variable     & $A_{0}$    & $A_{1}$   & $A_{2}$   & $A_{3}$   & $A_{4}$   &$\phi_{21}$ & $\phi_{31}$ & $\phi_{41}$
& $N$   &$D_m$ \\
  ID     & ($V$ mag)  & ($V$ mag)  &  ($V$ mag) & ($V$ mag)& ($V$ mag) & &  & & & \\
\hline
\multicolumn{11}{c}{RRab} \\
%&&&&&RRab&&&&&\\
\hline
V2 &16.952(2)& 0.407(3)& 0.170(3)& 0.136(3)& 0.089(3)& 3.741(23)& 7.698(30)&5.601(44)& 8&1.9\\
V3 &16.912(1)& 0.339(2)& 0.160(2)& 0.127(2)& 0.084(2)& 3.893(16)& 7.996(22)& 5.944(32)& 9&2.0\\
V4 &16.775(2)& 0.414(2)& 0.216(2)& 0.161(2)& 0.119(2)& 4.098(16)& 8.223(40)& 6.207(31)& 8&2.7\\
V6 &16.985(9)& 0.345(1)& 0.143(5)& 0.104(4)& 0.061(7)& 3.911(86)& 8.102(117)& 5.988(145)&9&1.9\\
V7 &16.872(2)& 0.262(2)& 0.128(2)& 0.072(3)& 0.032(2)& 4.181(27)& 8.468(40)& 6.714(79)& 7&1.9\\
V9  &16.965(3)& 0.321(3)& 0.141(3)& 0.118(4)& 0.073(3)& 3.733(33)& 7.829(42)& 5.660(70)& 9&4.8\\
V16 &16.896(2)& 0.250(3)& 0.003(3)& 0.121(3)& 0.003(3)& 4.067(34)& 8.530(48)& 6.720(100)& 9&1.6\\
V30 &16.916(3)& 0.286(4)& 0.142(4)& 0.096(4)& 0.061(4)& 4.088(39)& 8.430(60)& 6.465(90)& 9&1.5\\
V34 &16.991(2)& 0.362(3)& 0.157(3)& 0.107(4)& 0.071(3)& 4.003(28)& 8.212(37)& 6.318(55)& 7&2.2\\
V36 &16.884(2)& 0.418(5)& 0.198(6)& 0.155(4)& 0.102(6)& 3.774(40)& 7.840(63)& 5.770(69)& 9&2.2\\
V37 &17.009(2)& 0.382(4)& 0.177(4)& 0.125(4)& 0.078(4)& 3.907(30)& 8.116(45)& 6.087(78)& 9&1.1\\
V38 &16.907(6)& 0.347(8)& 0.176(8)& 0.138(9)& 0.088(8)& 3.984(70)& 8.054(88)& 5.974(13)& 9&2.6\\
V41 &16.980(2)& 0.350(2)& 0.164(3)& 0.108(3)& 0.067(3)& 3.925(20)& 8.058(32)& 6.157(48)& 9&1.5\\
V43 &16.969(3)& 0.314(4)& 0.157(4)& 0.124(4)& 0.070(4)& 3.994(35)& 8.379(47)& 6.345(75)& 9&2.1\\
V48 &16.922(2)& 0.301(3)& 0.159(3)& 0.112(3)& 0.066(3)& 3.900(26)& 7.982(39)& 6.119(58)& 9&3.6\\
V50 &16.984(2)& 0.190(3)& 0.071(3)& 0.038(3)& 0.018(3)& 4.150(54)& 8.707(94)& 6.805(173)& 7&9.7\\
\hline
\multicolumn{11}{c}{RRc} \\
%&&&&&RRc&&&&&\\
\hline
V11& 16.912(2) &0.269(3)& 0.048(2)&0.027(3)& 0.012 (2)& 4.600 (54)& 2.911  (94)& 1.713 (204)&4& \\
V23& 16.878(1) &0.258(2)& 0.053(2)&0.014(2)& 0.014 (2)& 4.664 (34)& 2.799 (122)& 1.263 (120)&4& \\
V26& 16.943(2) &0.178(3)& 0.035(3)&0.008(3)& 0.006 (3)& 4.640 (75)& 2.167 (336)& 1.398 (418)&4& \\
V53& 16.973(3) &0.279(3)& 0.060(4)&0.026(4)& 0.018 (4)& 4.853 (63)& 2.916 (140)& 1.579 (198)&4& \\
V56& 16.972(3) &0.276(5)& 0.069(5)&0.025(5)& 0.006 (5)& 4.679 (75)& 2.865 (190)& 0.771 (826)&4& \\
\hline
\end{tabular}
\end{center}
\end{table*}

\begin{table*}
\footnotesize
\begin{center}
\caption[] {\small Physical parameters of the \RRab and \RRc stars. The
numbers in parentheses indicate the uncertainty on the last
decimal places and have been calculated as described in the text.}
\label{fisicos}
 \begin{tabular}{lcccccccc}
\hline
Star&[Fe/H]$_{ZW}$ & [Fe/H]$_{UVES}$ &$M_V$ & log~$T_{\rm eff}$  &log$(L/{\rm
L_{\odot}})$ &$D$ (kpc)&
$M/{\rm M_{\odot}}$&$R/{\rm R_{\odot}}$\\
\hline
\multicolumn{9}{c}{RRab} \\
\hline
V2&-1.670(28)& -1.623(34)&  0.645(4)&  3.815(8)&  1.642(1)& 15.83(3)& 0.75(8)&
5.21(1)\\
V3&-1.608(21)& -1.542(24)&  0.624(3)&  3.810(8)&  1.650(1)& 15.69(2)& 0.69(6)&
5.38(1)\\
V4&-1.683(38)& -1.640(45)&  0.457(3)&  3.806(8)&  1.717(1)& 15.90(2)& 0.72(7)&
5.92(9)\\
V6&-1.569(110)& -1.493(124)& 0.574(3)&  3.810(18)& 1.670(1)& 16.61(2)& 0.70(15)&
5.50(1)\\
V7&-1.557(38)& -1.478(42)&  0.524(3)&  3.800(11)&  1.690(1)& 16.13(3)& 0.67(9)&
5.91(1)\\
V9&-1.800(39)& -1.800(51)&  0.623(5)&  3.807(10)&  1.651(2)& 16.09(4)& 0.70(9)&
5.47(12)\\
V16&-1.351(45)& -1.238(44)&  0.629(4)&  3.806(13)&  1.648(2)& 15.54(3)& 0.61(10)&
5.48(1)\\
V30&-1.388(56)& -1.280(56)&  0.588(6)&  3.808(12)&  1.665(2)& 15.99(4)& 0.65(9)&
5.52(1)\\
V34&-1.481(35)& -1.387(37)&  0.551(5)&  3.810(9)&  1.680(2)& 16.83(4)& 0.71(8)&
5.56(12)\\
V36&-1.588(59)& -1.517(67)&  0.625(7)&  3.816(10)& 1.650(3)& 15.48(5)& 0.73(9)&
5.24(2)\\
V37&-1.470(42)& -1.373(45)& 0.583(6)&  3.814(11)&  1.667(2)& 16.72(4)& 0.71(9)&
5.39(1)\\
V38&-1.492(83)& -1.399(89)&  0.649(12)&  3.813(7)&  1.640(5)& 15.47(8)& 0.68(9)&
5.24(3)\\
V41&-1.476(30)& -1.381(32)&  0.626(3)&  3.813(9)&  1.650(1)& 16.18(3)& 0.71(7)&
5.32(1)\\
V43&-1.336(44)& -1.222(42)&  0.616(6)&  3.812(11)&  1.654(2)& 16.16(4)& 0.64(8)&
5.36(1)\\
V48&-1.702(37)& -1.665(45)&  0.624(4)&  3.805(9)&  1.651(2)& 15.76(3)&  0.70(8)&
5.52(1)\\
V50&-1.220(88)$^{a}$& -1.102(78)$^{a}$&  0.618(4)&  3.806(21)&  1.653(2)& 16.26(3)& 0.61(15)&
5.50(1)\\
\hline
Weighted mean& $-$1.571(9)&$-$1.477(10)& 0.584(1)& 3.810(2)& 1.666(1)& 16.03(1)&
0.69(2)& 5.51(1)\\
$\sigma$&$\pm$0.137&$\pm$0.137&$\pm$0.052&$\pm$0.004&$\pm$0.020&$\pm$0.42&$\pm$0.04&
$\pm$0.21\\
\hline
%&&&RRc&&&&\\
\multicolumn{9}{c}{RRc} \\
\hline
V11&-1.66(17)&-1.62(20)&0.578(9)&3.865(1)&1.669(4)& 16.03(7)& 0.56(1)& 4.27(2)\\
V23&-1.48(21)&-1.38(22)&0.587(9)&3.869 (1)&1.665(44)& 15.71(7)& 0.59(1)& 4.17(2)\\
V26&-1.46(53)&-1.37(55)&0.650(14)&3.872(2)&1.640(5)& 15.72(10)& 0.62(2)& 4.00(3)\\
V53&-1.364(24)&-1.25(24)&0.565(18)&3.871(1)&1.674(7)& 16.58(14)& 0.61(1)& 4.18(4)\\
V56&-1.50(33)&-1.41(36)&0.618(22)&3.868(1)&1.653(9)& 16.17(17) &0.56(2)& 4.13(4)\\
\hline
Weighted mean&
$-$1.53(11)&$-$1.43(11)&0.593(3)&3.867(1)&1.662(2)&15.91(04)&0.58(1)&4.17(4)\\
$\sigma$&$\pm$0.11&$\pm$0.11&$\pm$0.010&$\pm$0.004&$\pm$0.035$\pm$0.09&$\pm$0.39&$\pm$0.27\\
\hline
\end{tabular}
\end{center}
\raggedright
\center{\quad $^{a}$Value not considered in the weighted mean.}
\end{table*}

The resulting physical parameters of the RR~Lyrae stars are
summarized in Table~\ref{fisicos}. The mean values given in the bottom of the table
are weighted by the statistical uncertainties. The iron abundance is given in the
scale of Zinn \& West (1984) and in the scale of Carretta et al. (2009). The
transformation between these two scales is of the form:

\begin{eqnarray}\label{UVES}
[\mathrm{Fe/H}]_{\mathrm{UVES}} &=& -0.413 +0.130\, [\mathrm{Fe/H}]_{ZW} \nonumber \\
  && -0.356\, [\mathrm{Fe/H}]_{ZW}^2.
\end{eqnarray}

Also listed are the corresponding distances.
Given the period, luminosity, and temperature for each RR~Lyrae star, its
mass and radius can be estimated from the equations: $\log~M/M_{\odot} =
16.907 - 1.47~ \log~P_F + 1.24~\log~(L/L_{\odot}) - 5.12~\log~T_{\rm eff}$ (van Albada
\& Baker 1971), and $L$=$4\pi R^2 \sigma T^4$ respectively.
The masses and radii given in Table~\ref{fisicos} are expressed in solar
units.

\subsection{Bailey diagram and Oosterhoff type}
\label{sec:Bailey}
\begin{figure}
\includegraphics[scale=0.75]{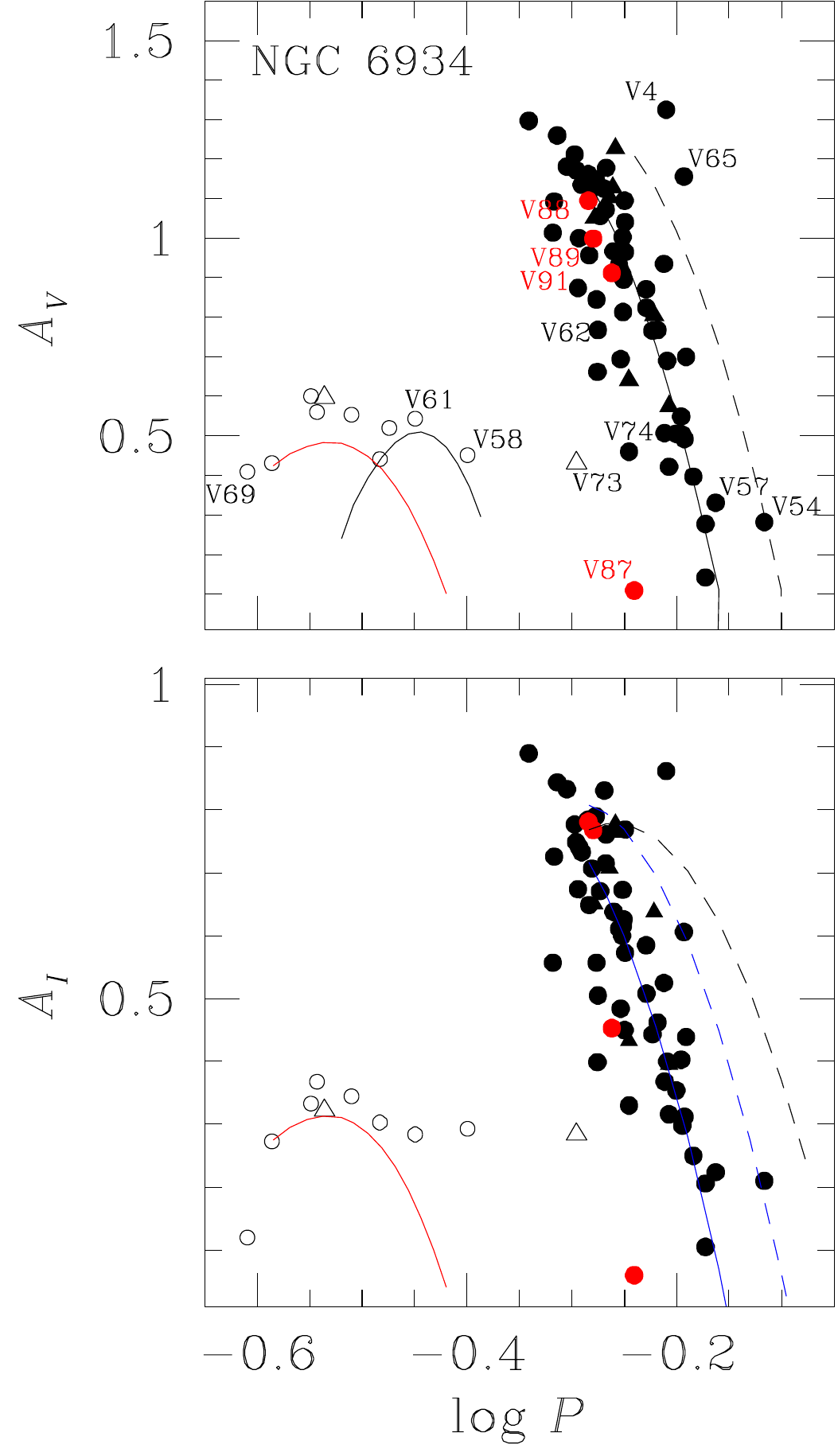}
\caption{Bailey diagram of NGC~6934 for $V$ and $I$ amplitudes. Filled and open
symbols represent RRab and RRc
stars, respectively. Triangles represent stars with Blazhko modulations. In the
top panel the continuous and segmented lines are the loci for evolved and unevolved
stars in M3 according to Cacciari et al. (2005). The black parabola was found by
Kunder et al. (2013b) from 14 OoII clusters. The red parabolas were calculated by
Arellano Ferro et al. (2015) from a sample of RRc stars in five OoI clusters and
avoiding Blazhko variables. In the bottom panel the black segmented locus was found
by Arellano Ferro et al.\ (2011; 2013) for the OoII clusters NGC~5024 and NGC~6333.
The blue loci are from Kunder et al. (2013). See \S~\ref{sec:Bailey} for details.}
    \label{Fig:Bailey}
\end{figure}

The period versus amplitude plane for RR~Lyrae stars is known as the Bailey
diagram. It is a useful tool in separating the RRab and the RRc stars as they occupy
markedly different regions in the diagram. The distribution of RRab stars also
offers insight on the Oosterhoff type of a globular cluster {\bf i.e. OoI or OoII,
the second group having RR Lyrae stars with a slightly longer period and being
systematically more metal poor}. The Bailey diagram of
M3 is usually used as a reference for OoI~clusters (see Fig.~4 of Cacciari et al.\
2005).
Fig.~\ref{Fig:Bailey} displays the corresponding distribution of the RR~Lyrae stars
in NGC~6934 with a good light curve coverage. The continuous and segmented
lines in
the top diagram represent, respectively, the mean distributions of non-evolved and
evolved stars in M3 according to Cacciari et al. (2005). For the RRc stars
distribution, the black parabola was calculated by Kunder
et al. (2013) from 14 OoII clusters while the red parabolas were calculated by
Arellano Ferro et al. (2015) for a sample of RRc stars in five OoI clusters and
avoiding Blazhko variables. It is clear from this figure that except for a few
outlier RRc stars (V49, V58 and V61), the
RRab and RRc stars in NGC~6934 follow the trends for OoI type cluster, which
identifies NGC~6934 as being of the type OoI.
For OoII clusters the RRab distribution is shifted toward longer periods and/or larger
amplitudes and do follow the segmented line in Fig.~\ref{Fig:Bailey} which is,
according to Cacciari et al. (2005), the locus of stars advanced in their evolution
toward the AGB; see for example the diagrams of the OoII cluster NGC~5024 (Arellano
Ferro et al.\ 2011 Fig.~7); NGC~6333 (Arellano Ferro et al.\ 2013, Fig.~17), and NGC
7099 (Kains et al. 2013, Fig. 10).

Arellano Ferro et al.\ (2011) also discussed the distribution of $A_I$
amplitudes in the
OoII cluster NGC~5024 and defined the locus shown as a black segmented line in the
bottom panel of Fig.~\ref{Fig:Bailey}. It has the equation:

\begin{eqnarray}\label{eq:AIOoIIab}
A_I &=& (-0.313 \pm 0.112) - (8.467 \pm 1.193)\, \log P \nonumber \\
 &&- (16.404 \pm 0.441)\, \log P^2.
\end{eqnarray}

The bottom panel of Fig.~\ref{Fig:Bailey} displays the distribution of the $I$
amplitudes of RRab and RRc stars in NGC~6934.
The blue loci are those calculated by Kunder et al. (2013) for the RRab stars
in OoI clusters (solid line) and in OoII
clusters (segmented line). Their OoI locus represents well the
distribution in NGC 6934. We note the difference in the locus of OoII
clusters proposed by Kunder et al. (2013) and the one observed by Arellano Ferro et
al. (2011; 2013) in NGC~5402 and NGC~6333 respectively (black segmented line).

Several outstanding stars are labeled in the top panel and they deserve a
dedicated discussion in Appendix A.

\section{The SX Phe stars}
\label{Sec:SXPHE}

\begin{figure}
\includegraphics[scale=0.45]{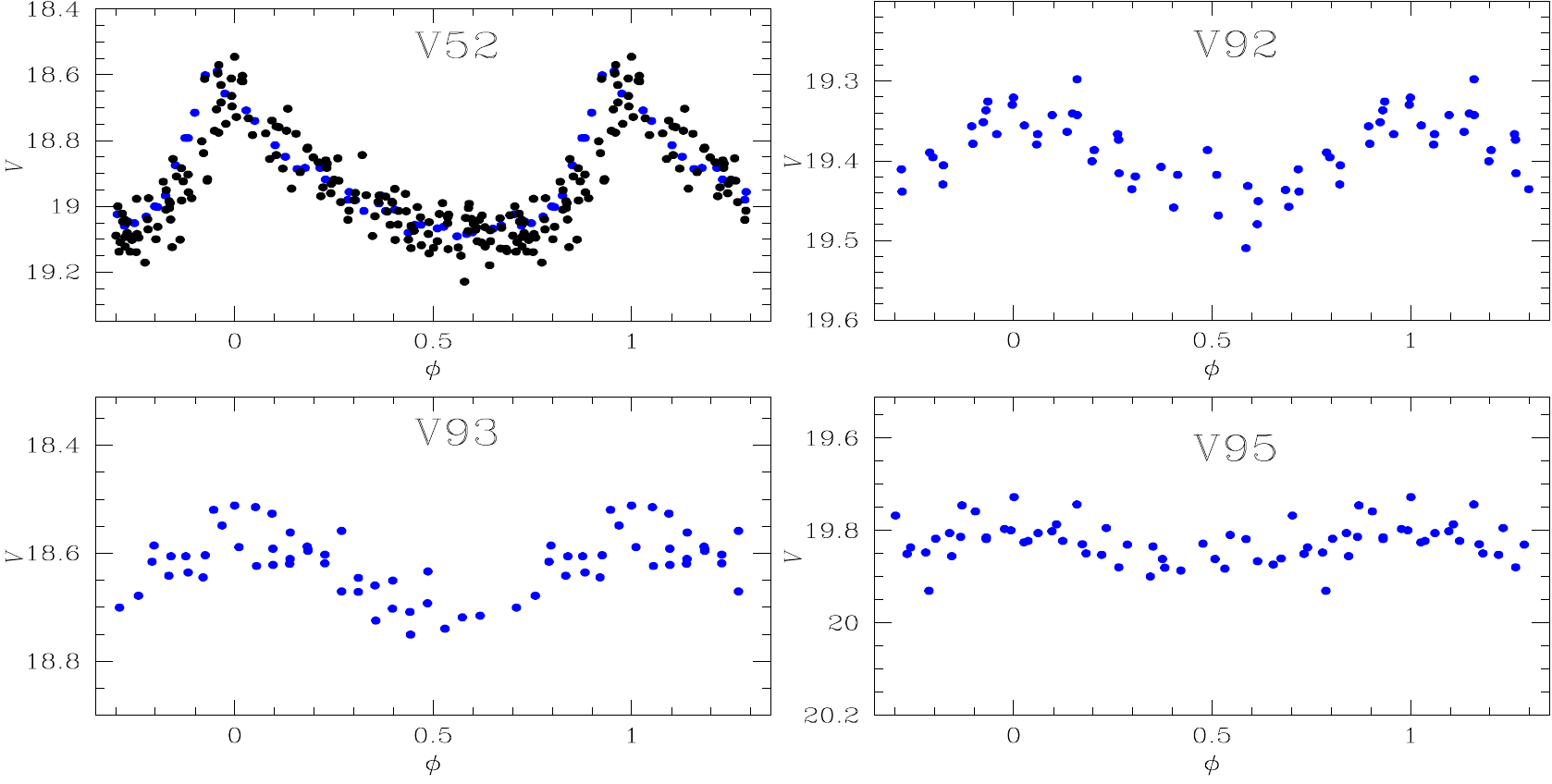}
\caption{$V$ light curves of the SX Phe star. V92, V93 and V95 are
new discoveries in this paper. These stars follow the P-L relation of SX Phe stars
and are therefore considered members of the cluster (see Fig. \ref{SXphe-PL}).}
    \label{SXphe}
\end{figure}

V52 was the only known SX Phe star in NGC~6934 prior to the present work.  A detailed exploration of the
residual images allowed us to discover three more, now labeled V92, V93 and V95. Their light curves are shown in Fig. \ref{SXphe}.
These four SX Phe will play a role in the determination of the distance to the cluster as explained in
$\S$ \ref{sec:DISTANCE}.

\section{Distance to NGC 6934 from its variable stars}
\label{sec:DISTANCE}

The distance to NGC 6934 can be estimated by a variety of approaches based on the variable stars.
Our first determinations come from the calculation of $M_V$ via the Fourier light curve decomposition
of the RRab and RRc stars. The average values of $M_V$ and distance are given in the bottom lines of Table
\ref{fisicos}. We found the distance values
16.03$\pm$0.42 kpc and 15.91$\pm$0.39 kpc respectively. Coming from independent
calibrations the results for the RRab and RRc
stars can be considered as two independent estimations.

Also for the RR Lyrae stars one can make use of the P-L relation for the $I$
magnitude derived by Catelan et al. (2004):

 \begin{equation}
M_I = 0.471 - 1.132 {\rm log} P + 0.205 {\rm log} Z,
\label{eqn:PL_RRI}
\end{equation}

\noindent
with ${\rm log}~Z =[M/H] -1.765$ and $[M/H] = \rm{[Fe/H]} - \rm {log} (0.638~f +
0.362)$ and log~f = [$\alpha$/Fe] (Salaris et al. 1993). For the sake of direct
comparison with the results of
Fourier decomposition we applied the above equations to the $I$ measurements of the 23 RR Lyraes in
Table \ref{fisicos} and found  a mean distance of 15.93$\pm$0.46 kpc, in excellent
agreement with the Fourier results.

Yet another approach to the calculation of the distance is from the P-L relation for
SX Phe stars (PLSX) which has been calibrated  by several
authors, notably Poretti et al.\ (2008) and McNamara (1997) for Galactic and
extragalactic $\delta$~Scuti and SX~Phe stars. In globular clusters the PLSX
has been studied by McNamara (2000) for $\omega$~Cen, Jeon et al.\ (2003) and Arellano
Ferro et al.\ (2011) for NGC~5024.
The calibrations of Arellano Ferro et al.\ (2011) for the fundamental mode of SX Phe
stars in NGC~5024 in the \textit{V} filter is of the form:
\begin{equation}
\label{PLV}
M_V = -2.916\, \log P - 0.898.
\end{equation}

This  calibration was used to calculate the distance to V52,V92 and V95. We adopted
the mean reddening $E(B-V)=0.10$ (Harris 1995), and found
the average 16.3$\pm$0.3 kpc which compares well with the independent
estimates from the RR Lyrae stars given above.
Alternatively, the P-L calibration of  Cohen \& Sarajedini (2012);
$M_v=-3.389\, \log P -1.640$ produces distances of 17.5 and 17.4 kpc respectively,
i.e. about 10\% larger, a trend already noted by Arellano Ferro et al. (2017).

\begin{figure}
\includegraphics[scale=0.45]{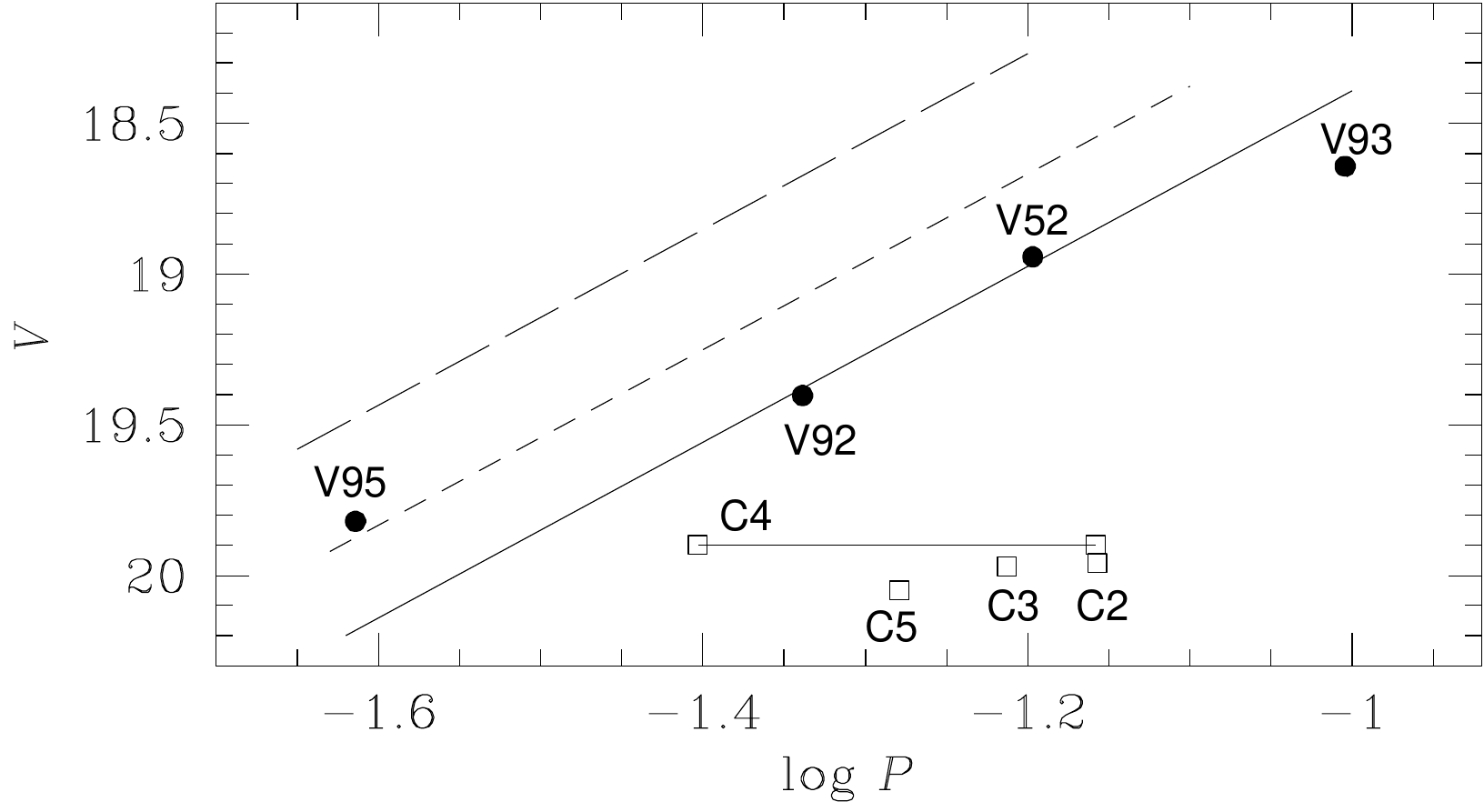}
\caption{The P-L relation of SX Phe Stars. Solid, and dashed lines are represent the
P-L calibration for the fundamental, first overtone and second overtone of Arellano
Ferro et al. (2011) scaled to the average distance of V52 and V92. V95 is a first
overtone pulsator, while V93 might be a little faint to be a cluster fundamental
pulsator. Open squares are likely SX Phe stars behind the cluster. Two periods are
plotted for the likely double mode C4. See Appendix A for a discussion on individual
stars.}
    \label{SXphe-PL}
\end{figure}

We then adopted the distance of 16.3 kpc and scale eq. \ref{PLV} for the fundamental
mode and plot the corresponding solid line in Fig. \ref{SXphe-PL}. The loci for the
first and
second overtone were drawn assuming the period rates $P_1/P_0 = 0.783$ and $P_2/P_0 =
0.571$ (see Santolamazza et al. 2001 or Jeon et al. 2003; Poretti et al. 2005). It
seems clear from this figure that V52, V92 and V95 are cluster members
and that the later pulsates in the first overtone.
For V93 the distance is about 18.2 kpc, a bit too large to be a cluster member. Other
candidate SX Phe stars
labeled in the plot shall be discussed in the Appendix A.

\begin{table*}
\footnotesize
\caption{Time-series \textit{V} and \textit{I} photometry for all the confirmed variables in our
field of view. The standard \Mstd and
instrumental \mins magnitudes are listed in columns 4 and~5,
respectively, corresponding to the variable stars in column~1. Filter and epoch of
mid-exposure are listed in columns 2 and 3, respectively. The uncertainty on
\mins is listed in column~6, which also corresponds to the
uncertainty on \Mstd. For completeness, we also list the
reference and differential fluxes \fref
and \fdif and the scale factor \lowercase{\textit{p}}
in columns 7, 9, and~11, along with the uncertainties \sref
and \sdiff in columns 8 and~10. This is an extract from
the full table, which is available with the electronic version of the article.
         }
\centering
\begin{tabular}{ccccccccccc}
\hline
Variable &Filter & HJD & $M_{\mbox{\scriptsize std}}$ &
$m_{\mbox{\scriptsize ins}}$
& $\sigma_{m}$ & $f_{\mbox{\scriptsize ref}}$ & $\sigma_{\mbox{\scriptsize ref}}$ &
$f_{\mbox{\scriptsize diff}}$ &
$\sigma_{\mbox{\scriptsize diff}}$ & $p$ \\
Star ID  &    & (d) & (mag)     & (mag)   & (mag) & (ADU s$^{-1}$) &(ADU s$^{-1}$)
               &(ADU s$^{-1}$)  &(ADU s$^{-1}$)    & \\
\hline
V1 &$V$ &2455779.37293& 16.996& 18.125 & 0.005 &  738.417 & 2.078 & $-$195.326&2.881&1.1081\\
V1 &$V$ &2455779.37758& 17.010& 18.139 & 0.005 &  738.417 & 2.078 & $-$208.125&2.978&1.1349\\
\vdots   &  \vdots  & \vdots & \vdots & \vdots & \vdots   & \vdots & \vdots  & \vdots&\vdots \\
V1 &$I$ &2455779.36627& 16.307& 17.357& 0.007&   1195.901&  4.821& $-$55.859&  7.876& 1.0051\\
V1 &$I$ &2455779.37043& 16.319& 17.369& 0.008&   1195.901&  4.821& $-$67.854&  8.632& 1.0056\\
\vdots   &  \vdots  & \vdots & \vdots & \vdots & \vdots   & \vdots & \vdots  & \vdots&\vdots \\
V2 &$V$&2455779.37293 &17.370& 18.493& 0.007&    392.308&  2.364& $+$9.324 &2.752& 1.1081\\
V2 &$V$&2455779.37758 &17.379& 18.502& 0.007&    392.308&  2.364& $+$5.722 &2.801& 1.1349\\
\vdots   &  \vdots  & \vdots & \vdots & \vdots & \vdots   & \vdots & \vdots  & \vdots&\vdots \\
V2 &$I$&2455779.36627& 16.699&  17.746&  0.010&  805.480&   4.515 &$-$8.408 & 7.296& 1.0051\\
V2 &$I$&2455779.37043& 16.677&  17.724&  0.010&  805.480&   4.515 &$+$7.914 & 7.929& 1.0056\\
\vdots   &  \vdots  & \vdots & \vdots & \vdots & \vdots   & \vdots & \vdots  & \vdots&\vdots \\
\hline
\end{tabular}
\label{tab:vi_phot}
\end{table*}

A last approach we used to estimate the cluster distance is using the variables
near the tip of the RGB (TRGB).
This method, originally developed to estimate distances to nearby galaxies (Lee et
al. 1993) has already
been applied by our group for the distance estimates of other clusters e.g. Arellano
Ferro et al. (2015) for
NGC 6229 and Arellano Ferro et al. (2016b) for M5. In the former case the method
was described in detail.  In brief, the idea is to use the bolometric magnitude
of the tip of the RGB as an indicator. We use the calibration of Salaris \& Cassisi
(1997):

\begin{equation}
\label{TRGB}
M_{bol}^{tip} = -3.949\, -0.178\, [M/H] + 0.008\, [M/H]^2,
\end{equation}

\noindent
where $[M/H] = \rm{[Fe/H]} - \rm {log} (0.638~f + 0.362)$ and log~f = [$\alpha$/Fe]
(Salaris et al. 1993). However one should take into account the fact that the true
TRGB might
be a bit brighter than the brightest observed stars, as argued by Viaux et al. (2013)
in their analysis of M5, under the arguments that the neutrino magnetic dipole moment
enhances the plasma decay process, postpones helium ignition in low-mass stars, and
therefore extends the red giant branch (RGB) in globular clusters. According to these
authors the TRGB is between 0.05 and 0.16 mag brighter than the brightest stars on the
RGB. Therefore the
magnitudes of the two brightest RGB stars in NGC 6934, V15 and V86
would have to be corrected by at least the above quantities to bring them to the TRGB.
Applying the corrections 0.05 and 0.16 we find distances of 16.6 kpc and
15.7 kpc respectively. If on the other hand we accept, from the results for the RR
Lyrae and SX Phe discussed above, that the distance to NGC 6934 is between 15.9 and
16.1 kpc, then the correction for the TRGB should be about 0.12 for NGC 6934.

Table \ref{Sum_Dis} summarizes the values of the distance obtained by the
approaches described above.

\begin{table*}[t]
\footnotesize
\caption{Distance to NGC 6934 by different approaches.}
\centering
\begin{tabular}{lcc}
  \hline
  Approach & Calibration & Distance\\
   &   & (kpc)\\
  \hline
  Fourier light curve decomposition of the RRab & Kov\'acs \& Walker (2001) & 16.03$\pm$0.42\\
  Fourier light curve decomposition of the RRc & Kov\'acs (1998) & 15.91$\pm$0.39\\
  RR Lyrae $I$-magnitude P-L relation& Catelan et al. (2004) & 15.93$\pm$0.46\\
  SX Phe P-L relation & Arellano Ferro et al. (2011) & 16.3$\pm$0.3\\
  Bolometric magnitude of the TRGB & Salaris \& Cassini (1997) & 15.9-16.1$^a$ \\
  \hline
\end{tabular}
\label{Sum_Dis}
%\raggedright
\center{\quad $^{a}$Exact value is subject to the correction of the true TRGB. The
given range is compatible with \\
a correction of about 0.12 mag (see $\S$ \ref{sec:DISTANCE} for details).}
\end{table*}

\section{The HB of NGC~6934 and Probable evolved stars}
\label{sec:HB}

The instability strip at the level of the HB is populated by RR Lyrae stars evolving
both to the blue and to the red. According to a scheme described by Caputo et al.
(1978) and sustained on theoretical grounds; depending on the mass of the pre-HB
star, the ZAHB evolutionary track starts in first overtone (FO) instability strip as
an RRc, in the Fundamental mode (F) instability strip as RRab, or in the inter-order
"either-or" region, in which the initial pulsating mode depends on the pre-HB phase
related to the onset of CNO. This mechanism may produce an either-or region populated
by both RRc and RRab stars or a distribution of clearly separated modes at the red
edge of the first overtone instability strip. In the later case the average period of
the RRab stars would be larger (like in OoII clusters) than in the former (the OoI
case). One may expect, under this scheme, that the distribution of RRab and
RRc stars in the instability strip tends to present a clear segregation of modes in
the OoII and not so in the OoI clusters.

\begin{figure*}
\begin{center}
\includegraphics[scale=0.7]{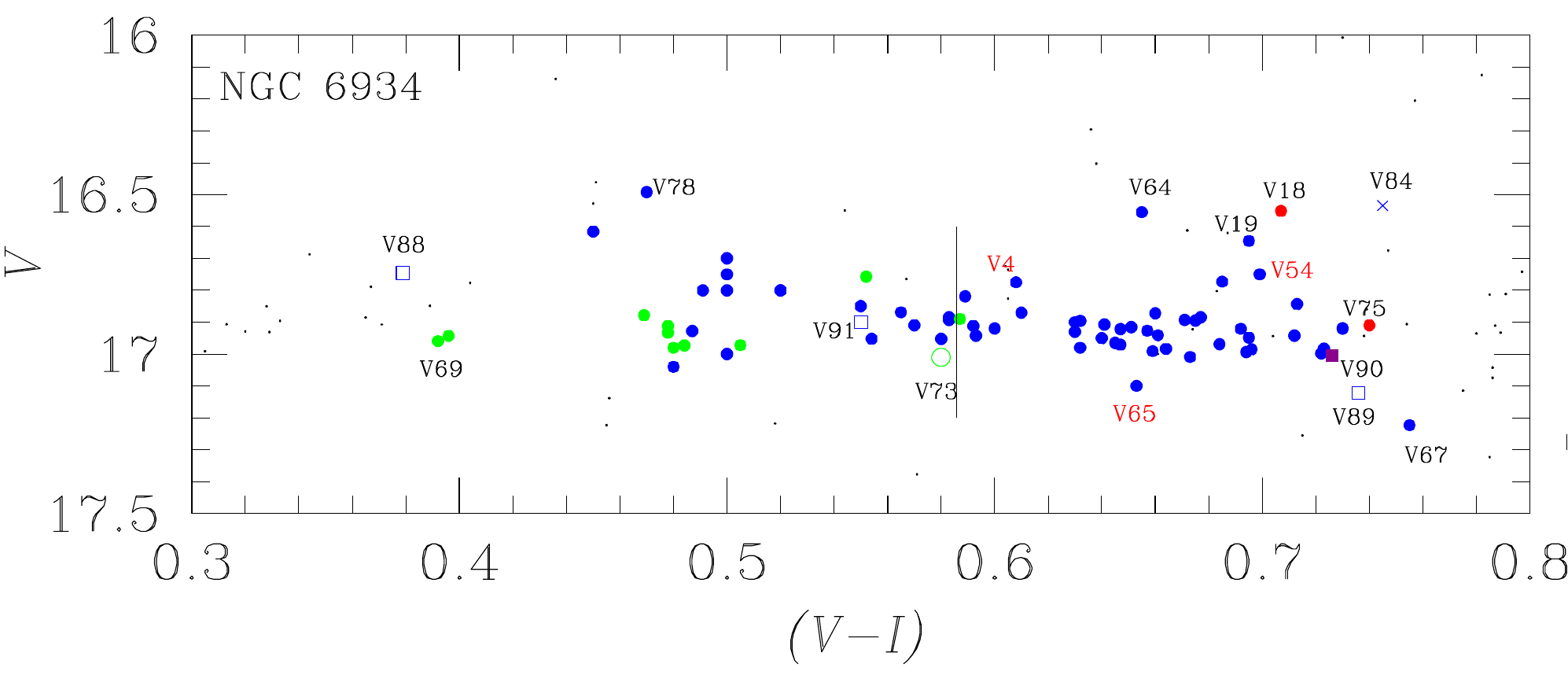}
\caption{The distribution of RR Lyrae stars is shown in detail in this blow-up of the
HB in NGC 6934. The colour code is as in Fig. \ref{CMD}. The vertical line
indicates the first overtone red edge as estimated empirically (Arellano Ferro et al.
2016b) and helps realizing that in NGC 6934 the intermode or "either-or" regions
is shared by RRab and RRc stars. See $\S$ \ref{sec:HB} for further discussion.}
\label{HB}
\end{center}
\end{figure*}

The distribution of RRc and RRab in the HB of several OoI and OoII clusters has been
addressed by Arellano Ferro et al. (2015, 2016a). In summary, neat RRc-RRab
segregation has been observed in all OoII clusters studied by these authors, named
NGC~288, NGC~1904, NGC~4590, NGC~5024, NGC~5053,  NGC~5466, NGC~6333, NGC~7099. On the
other hand, of the studied OoI clusters NGC~3201, NGC~5904, NGC~6229, NGC~6362 and
NGC~6934, NGC 6229 and NGC 6362 present a clean segregation of the modes whereas in
the others the either-or region is populated by both RRc and RRab stars. The case of
NGC~6934 is
illustrated in Fig. \ref{HB} where we have drawn a vertical black line at the border
between the retribution of RRc and RRab stars as observed in several clusters.
This has been interpreted by Arellano Ferro et al. (2016b) as the empirical
red edge of the first overtone instability strip (RFO) and estimated it at $(V-I)_0
\sim 0.46$.
This border was reddened by $E(B-V)=0.1$ and assuming $E(V-I)= 1.259 E(B-V)$
(Schlegel et al. 1998; Table 6) resulting at $(V-I)= 0.586$. Clearly in NGC 6934,
while the RRc stars fall to the blue of the edge, the RRc and RRab stars share the
either-or region.

\begin{figure}
\includegraphics[scale=0.48]{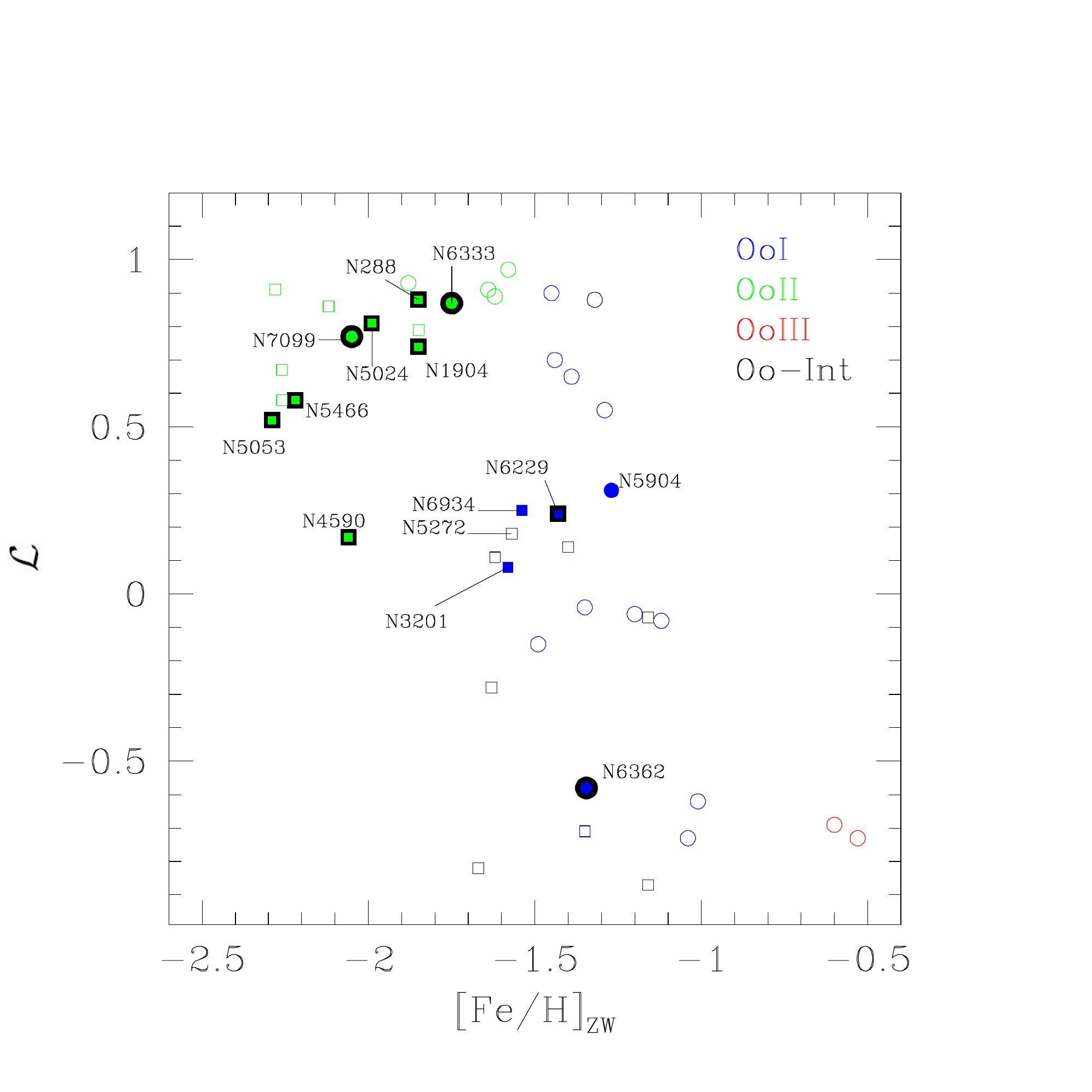}
\caption{Distribution of OoI and OoII clusters in the [Fe/H]$-\cal L$ plane. Filled
symbols are used for cluster whose HB's have been studied in detail and those with a
black rimmed show a clear RRc-RRab segregation, i.e. all studied OoII clusters but
only some OoI. Circles and squares are used for inner and outer halo clusters
respectively.}
    \label{L-Fe}
\end{figure}

Fig. \ref{L-Fe} illustrates the distribution of OoI and OoII clusters in the
[Fe/H]$-\cal L$ plane, where $\cal L$ is the Lee-Zinn parameter defined as
$(B-R)/(B+V+R)$ where $B,V,R$ refer to the number of stars to the blue, inside, and to
the red of the IS. Large values of $\cal L$ indicate HB's with long blue tails while
very negative values correspond to clusters with no blue tail and rather red clumps.
All OoII have blue tails, with the possible exception NGC 4590, whereas the OoI
clusters display an assortion of HB structures. In Fig.\ref{L-Fe} filled symbols
represent those clusters studied by our group and those symbols with a black rim are
for the clusters with a clear RRc-RRab segregation, i.e. all OoII clusters and some,
NGC 6229 and NGC 6362, among the OoI clusters. No trend is seen for inner and outer halo
clusters neither in terms of $\cal L$ nor in the RRc-RRab segregation.

These observational results, if interpreted according to the scenario proposed by
Caputo et al. (1978), imply that in fact RR Lyrae stars in Oo II clusters are
evolved from less massive stars starting their evolution across the instability strip
on the bluer part of the ZAHB as first overtone pulsators, leading to either-or
regions populated exclusively by RRc stars and with RRab stars averaging longer
periods. On the other hand, in some OoI clusters, RR Lyrae star may start their
evolution as in OoII clusters (e.g. NGC 6229 and NGC 6362) or if they have a
population of more massive stars starting in the redder part of the
ZAHB, then produce an either-or region populated by both pulsating modes (e.g. NGC
3201, NGC 5904, NGC 6934).
What determines these two circumstances in OoI clusters is not
clear, although mass loss in the RGB is most likely the answer. Pre-ZAHB evolutionary
models calculated with a range of mass-loss efficiencies (Silva-Aguirre et al. 2008)
show that if mass loss is efficient, low mass stars will populate preferentially the
bluer part of the ZAHB and viceversa. Hence the Oosterhoff type of the cluster
must be determined by the mass loss rates being driven in a given system, which in
turn may be connected with the primordial chemistry of a particular cluster. One
further complication to adopt a single scenario is the growing evidence of the
presence of more that one stellar population (e.g. Gratton et al. 2004, Milone et al.
2009, Carretta et al. 2010) at least in some clusters. This will
impact on the mass distribution on the ZAHB and hence on the subsequent distribution
of stars in the CMD. Mass distribution on the ZAHB have been studied in detailed for
NGC 5272 (M3), which is considered the prototype of OoI type clusters (Rood \&
Crocker 1989; Valcarce \& Catelan 2008) showing that ZAHB are distributed on both
sides of the RFO and which suggests that the either-or regions should be shared by
RRc and RRab stars. That this is the case can be seen in the CMD of Valcarce et al.
(2008) (Fig. 2). Similar studies in other clusters would be
very enlightening in the understanding of the observed stellar distributions on the
HB of both Oosterhoff type of clusters.

To identify, among a population of cluster RR Lyrae stars, those that may be truly
advanced in their evolution towards the AGB is an observational challenge. There are
some indicators that, in favorable circumstances, can help identifying evolved stars,
e.g. the distribution of RRab stars in the Bailey diagram, as discussed by Cacciari
et al. (2005) for M3, or the secular large positive period changes, like those cases
identified in M5 by Arellano Ferro et al. (2016a), although extreme values of $\dot P$
can be achieved by the very rapid evolution in pre-ZAHB according to Silva-Aguirre et
al. (2008). In the present case of NGC 6934, in
Fig. \ref{Fig:Bailey} we note that stars V4, V54 and V65 fall along the evolved star
sequence. Their position on the HB (labeled with
red numbers in Fig. \ref{HB}) show V4 and V54 among the brightest RRab as expected for
evolved stars. On the contrary V65 is among the faintest. A crucial test for the
evolutionary stage of these stars would be to explore their secular period behaviour,
$\dot P$.
If truly advanced in their evolution towards the AGB, large positive values of $\dot
P$ are expected. Unfortunately to estimate $\dot P$, data for a large time base are
necessary. In the case of NGC 6934 previous photometric studies include that of KOS01,
Sawyer-Hogg (1938) and Sawyer-Hogg \& Wehlau (1980). While most early light curves
are no longer available, we shall try to retrieve times of maximum light or phase
information from published material. The results of that effort will be reported
elsewhere.

\section{Summary of results}
\label{sec:Summ}

We have performed a new CCD photometric study of the globular cluster NGC 6934 and
have analyzed the variable stars individually with the aim to
confirm their classification and to estimate their physical parameters, particularly
the absolute magnitudes and [Fe/H] which in turn lead to the mean values of the
distance and metallurgist of the parental globular cluster. For the RR
Lyrae stars we
performed the Fourier decomposition of their light curves to estimate the mean iron
abundance and distance [Fe/H]$_{ZW}$=--1.57$\pm$0.13
([Fe/H]$_{UVES}$=--1.48$\pm$0.14) and distance 16.03$\pm$0.42 kpc
from the RRab stars, and [Fe/H]$_{ZW}$=--1.53$\pm$0.11
([Fe/H]$_{UVES}$=--1.43$\pm$0.11) and 15.91$\pm$0.39 kpc from the RRc stars, which
coming from independent calibrations can be considered as two independent estimations.

Independent distances to the cluster were also calculated via the $I$ P-L of RR Lyrae
(Catelan et al. 2004); the P-L relation of SX Phe (Arellano Ferro et al. 2011) and
from the bolometric magnitude estimation of the tip of the RG branch (Salaris \&
Cassisi 1997) and found the values;  15.9$\pm$0.5 kpc, 16.3$\pm$0.3 and 15.7-16.6
kpc
respectively.

We detected 12 new variables; 4 RR Lyrae, 3 SX Phe, 2 Pop II cepheids or CWB, and
three semiregural or SR stars. Also one RR Lyrae and four SX Phe were detected that
are probably not members
of the cluster for which further exploration is recommended.

We found that inter-order, either-or, region on the HB of NGC 6934 is occupied by both
RRc and RRab stars,
a characteristic shared with the OoI type clusters NGC 3201, NGC 5272 (M3) and NGC
5904 (M5) and at odds with all OoII type clusters we have studied and the two OoI
clusters
NGC 6229 and NGC 6362. We have speculated that this property is a consequence of the
mass loss rates involved during the He flashes and hence the resultant mass
distribution on the
ZAHB. Further work in the observational-theoretical interface will most likely
contribute to the understanding the evolutionary processes and chemical conditions
behind the observed stellar distributions on the HB.

\vskip 2.0cm

\noindent
We are indebted to Dr.~Daniel Bramich for allowing us the use DanDIA and
for enriching our work with very constructive comments.
AAF acknowledges the support from DGAPA-UNAM grant through project IN104917.
We have made an extensive use of the SIMBAD and ADS services, for which we are
thankful.

\appendix

\section{Appendix}
\label{Appendix}

\subsection{Comments on individual stars}

In this section we only discuss those stars that deserve particular comments.

V10, V17, V22, V27, V28, V29, V75, V76, V92. These are all out of field in setting B
and
their light curves are only partially covered from our setting A data.

V12. This star in our images appears blended with a very close neighbour of similar
brightness and we have not being able to resolve its light curve. The star is not
included in our analysis.

V18. KOS01 found a period of 0.956070 d for this star, which is much too long even for
an
RRab. In fact this period places the star too far to the right in the Bailey diagram
of Fig. \ref{Fig:Bailey}. With that period our light curve is incomplete. An
alternative period of 0.484816 d produces a sinusoidal light curve but then the star
falls between the RRab and the RRc loci in Fig. \ref{Fig:Bailey}. The star is
about half a magnitude brighter than the average HB. In our opinion this is not an RR
Lyrae star but a Pop II Cepheid or W Virginis (CWB). The light curve in Fig
\ref{RRL_A} and in KOS01, exhibits a bump on the rising branch, typical of Pop II
Cepheids. Applying the $V$ P-L relation for Pop II Cepheids from Pritzl et al. (2003)
we find the distance of 17.0$\pm$0.4 kpc, which, given the uncertainties, is only
slightly larger than the cluster distance values found in $\S$ \ref{sec:DISTANCE}.

V13, V20, V21, V35. They all display very prominent Blazhko modulations in
both $V$ and $I$.

V51. This is a bad blend but from image blinking the variable is the fainter star
to the east of the pair.

V57. This is an RRab whose low amplitude and large period place it in the bottom
of the RRab distribution in the Bailey diagram (Fig. \ref{Fig:Bailey}). However the
star appears more than one magnitude above the HB, hence the star is likely not a
cluster member but a nearer field variable. In fact the light curve Fourier
decomposition suggests a distance of only 9.6 kpc, compared with the $\sim$16 kpc
of the cluster. Alternatively the star may be a short-period W Virginis star or
CWB that might in fact not be a cluster member.

V61. This is not an RRab but an RRc star since its period and amplitudes in $V$ and
$I$ place the star among the RRc stars in the Bailey diagram.

V62, V63. V62 is blended with two fainter stars and probably contaminated with V63.
On the other hand V63 is relatively isolated but again a little
  contamination from V62 hence the light curve is somewhat noisy.
Given the scale of our images and the seeing
conditions may explain the uncomfortable scatter in these two stars. We note however
the peculiar position of V62 in the CMD. We therefore refrain from making comments on
the variable nature of these two stars.

V66. This star is very close to the cluster center and badly blended in our images
with another star of similar brightness. A careful analysis of our collection of
differential images do not show any convincing variability of this star. We were
unable to isolate its light curve in both our settings and then we refrain from
further analysis. The star is reported as a large amplitude RRab star by KOS01 but
these authors were not able to estimate the amplitude nor the mean magnitude. This
star deserves a fresh monitoring.

V68. This star is blended with at least another two stars in our images and it was
impossible to isolate its flux. As a consequence the shape and
amplitude of the light curve displayed in Fig. \ref{RRL_A} is largely distorted by the
contamination of the neighbours. We did not use this star for any purpose.

\begin{figure}
\begin{center}
\includegraphics[scale=0.35]{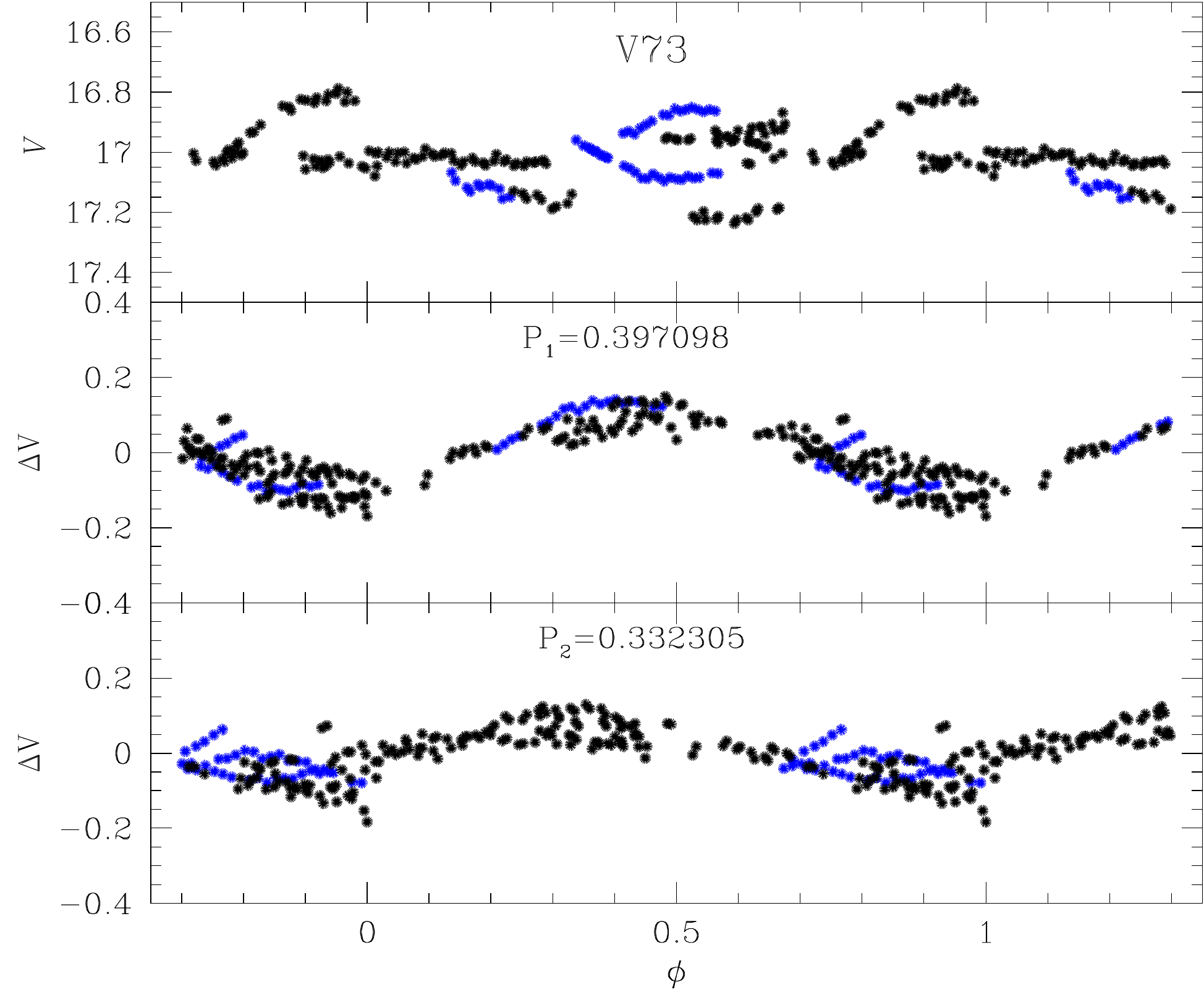}
\caption{In the top panel the $V$ Light curve of V73 phased with the period 0.506209
proposed by KOS01. Lower two panels show the disentangled contribution to the light
curve of two frequencies found in our analysis.}
\label{LCV73}
\end{center}
\end{figure}

V73. This star is reported as an RRab in the CVSGC. The light curve of KOS01 shows
large
amplitude modulations and our light curve cannot be phased with the period of
0.506209d given by KOS01. We have found two active periods in the power spectrum of
our data. Although our
data are not ideal for an accurate determination
of the involved periods, two structures in the power spectrum are evident. Once
these are prewhited some signal might remain but we have essentially reached the
noise level. The periods are roughly 0.3971d and 0.3323d. The period
ratio is
0.83 which is a bit off the canonical value 0.74 for radial mode ratio, however, we
recognize that the periods and period ratio are not very accurate and need to be
confirmed with more appropriate data. The double mode nature of the star is
however clear. In Fig. \ref{LCV73}, the top panel shows how
the period 0.506209 reported by KOS01 fails at proper phasing of our data. The bottom
panels display the disentangled contribution to the light curve of each the two
modes. The remaining power in the
spectrum after the removal of the two modes and the visible
scatter remaining in the light curves are a consequence of the limited data in our
possession.

%\begin{figure}
%\begin{center}
%\includegraphics[scale=0.4]{V73_spectra.pdf}
%\caption{Power spectra of V73 showing the two active modes.}
%\label{spectra}
%\end{center}
%\end{figure}

V77. We have not been able to locate this star, classified by KOS01 as a log period variable
of the L type. The coordinates provided by KOS01 for V77 are those of V79 and then
the published chart of V77 does not help.

V85. We confirm the variability of this star with a period and position on the CMD
of Fig. \ref{CMD} very similar to the reported by KOS01. The star might well be a W
Virginis star behind the cluster. The star is not identified in the finding chart of
Fig. \ref{chart} because it is not contained in our setting A. All our data for this
star come from setting B.

V86. We found that given its coordinates the finding chart in KOS01 points to the
wrong star. The correct star is identified in out chart in Fig. \ref{chart}.

V87. The variability of this star is very clear and its period and light curve shape
first suggest that it is an RRab star. However its position in the CMD about 2
magnitudes above the HB and its odd position on the Bailey diagram with very small
amplitude make us believe that either we are dealing with W Virginis star or CWB or
rather a foreground RRab. The later options is not likely given the fact that the
star is about 0.5 mag too red. Thus we have tentatively classified it as a CWB.

V88, V89, V91. These RRab star are located the core of NGC 6934, very near the center
and
they escaped detection in previous studies. Our light curves, although sometimes
noisy due to contamination of nearby neighbours, are clear in shape and period
and their amplitudes are as expected for RRab of their corresponding period (Fig.
\ref{Fig:Bailey}), however, their position on the CMD may not be accurate (Fig.
\ref{HB}).

V90. Although if falls well within the RRab region on the HB, its period of 1.05 d
makes us believe that we are dealing with a CWB star.

V92, V93. These are two clear SX Phe star. While V92 (like V52) falls right on the
fundamental P-L relation for SX Phe stars, V93 is a bit too faint, suggesting that
the star might be behind the cluster as indicated by the SX Phe P-L relation (see
$\S$ \ref{sec:DISTANCE}).

V94. This is a RRab star more than a magnitude brighter than the mean HB. It is likely
a foreground star.

\begin{figure}
\begin{center}
\includegraphics[scale=0.40]{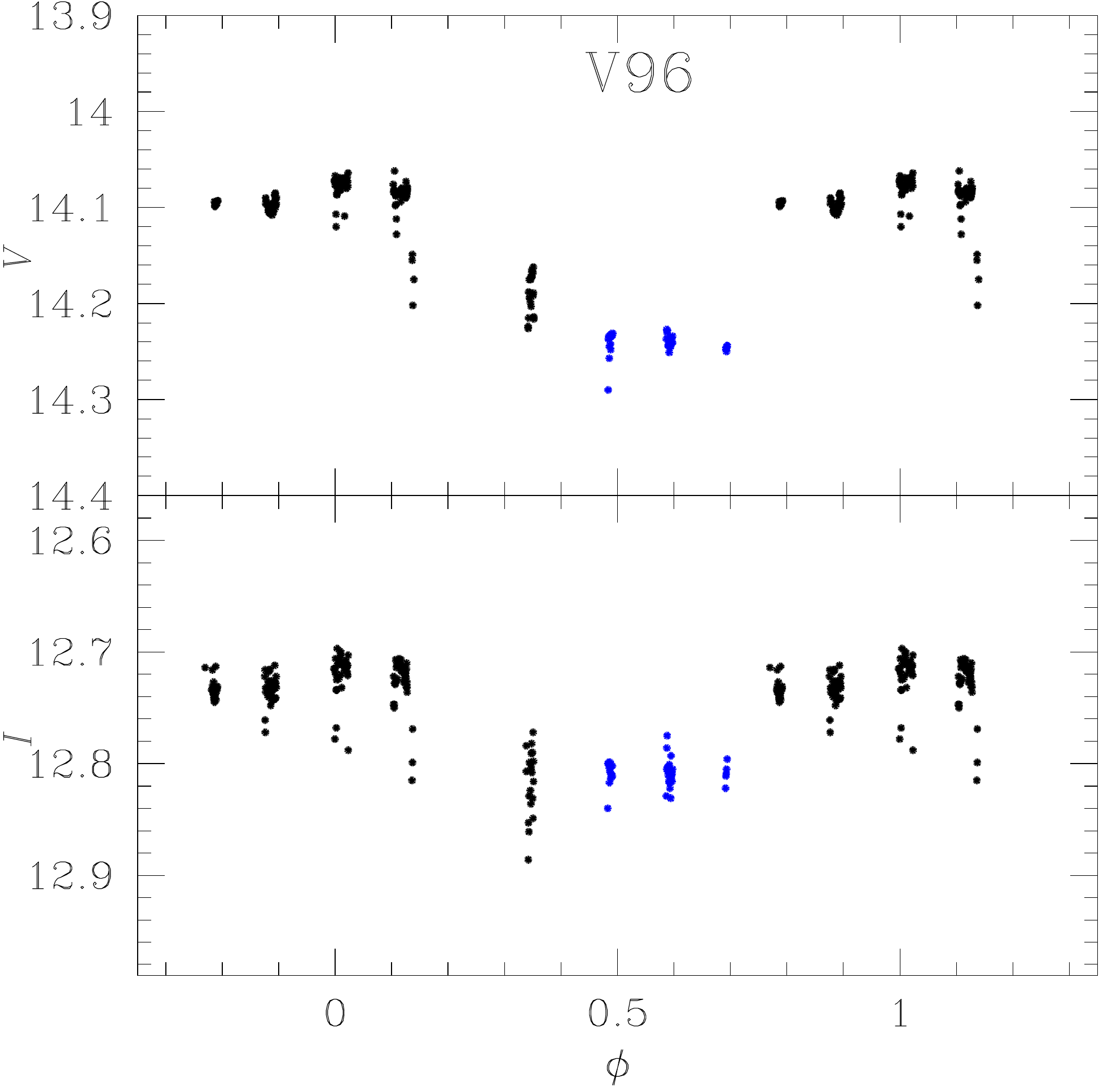}
\caption{V96 phased with a period of 9.54d.}
\label{V96}
\end{center}
\end{figure}

V96-V98. The variability of this stars is suggested in a HJD vs. $V$ plot of (Fig.
\ref{SR_HJD}) and was confirmed by inspecting the residual images. Our data are not
fit for a period determination however for the case of V96 a period of 9.54 d seems a
reasonable first estimate as suggested by the
phased $V,I$ light curves of Fig. \ref{V96}.

C1, C2, C3, C4, C5. These five variables may require confirmation from better data,
hence we
have not assigned a variable number. C1 shows a clear maximum light but a flat bottom
(Fig. \ref{Cand}). Its position in the CMD to the red of the RGB is unexpected and we
have not been able to suggest a classification. C2-C5 are most likely SX Phe
stars however, they are too faint for our detection and classification be
sufficiently solid. C4 is a likely double mode star with periods 0.03953 d and 0.06951
d. They do not follow the SX Phe P-L relation in Fig. \ref{SXphe-PL}, scaled to the
distance estimated from V52, V92 and V93, thus while SX Phe they may be, they are
probably not
cluster members.

\end{document}